\documentclass[twocolumn,twocolappendix,iop,numberedappendix,appendixfloats]{openjournal}

\usepackage[utf8]{inputenc}

\usepackage[british]{babel}

\usepackage{amsmath}
\usepackage{amssymb}
\usepackage{graphicx}
\usepackage{bm}
\usepackage{natbib}
\usepackage[colorlinks,allcolors=blue]{hyperref}

\usepackage[font=small,labelfont=bf]{caption}
\usepackage{subcaption}
\usepackage[dvipsnames]{xcolor}
\usepackage{verbatim}
\usepackage{lineno}

\renewcommand{\vec}{\boldsymbol}

\newcommand{\beq}{\begin{eqnarray}}
\newcommand{\eeq}{\end{eqnarray}}

\newcommand{\bit}{\begin{itemize}}
\newcommand{\eit}{\end{itemize}}
\newcommand{\ben}{\begin{enumerate}}
\newcommand{\een}{\end{enumerate}}

\newcommand{\tpcf}[1]{\xi_{\rm #1}}
\newcommand{\tpcftwo}[2]{\xi_{\rm #1}^{\rm #2}}
\newcommand{\evec}[1]{\vec{e}_{\rm #1}}

\newcommand{\halotools}{{\tt halotools}}

\newcommand{\dd}{\rm d}
\newcommand{\mean}[1]{\langle #1 \rangle}
\newcommand{\meantwo}[2]{\langle #1 \vert #2 \rangle}

\begin{document}

\journalinfo{The Open Journal of Astrophysics}
\submitted{Accepted June 2, 2024}

\shorttitle{An Empirical Model for Intrinsic Alignments}
\shortauthors{N. Van Alfen et al.}

\title{An Empirical Model for Intrinsic Alignments: Insights from Cosmological Simulations}

\author{Nicholas Van Alfen$^{1}$}
\author{Duncan Campbell$^{2}$}
\author{Jonathan Blazek$^{1}$}
\author{C. Danielle Leonard$^{3}$}
\author{Francois Lanusse$^{4}$}
\author{Andrew Hearin$^{5}$}
\author{Rachel Mandelbaum$^{2}$}
\author{the LSST Dark Energy Science Collaboration}

\affiliation{$^{1}$Department of Physics, Northeastern University, Boston, MA 02115, USA}
\affiliation{$^{2}$McWilliams Center for Cosmology, Department of Physics, Carnegie Mellon University, Pittsburgh, PA 15213, USA}
\affiliation{$^{3}$School of Mathematics, Statistics and Physics, Newcastle University, Newcastle upon Tyne, NE1 7RU, UK}
\affiliation{$^{4}$Université Paris-Saclay, Université Paris Cité, CEA, CNRS, AIM, 91191, Gif-sur-Yvette, France}
\affiliation{$^{5}$Argonne National Laboratory, Lemont, IL 60439, USA}
\thanks{$^\star$ E-mail: \nolinkurl{vanalfen.n@northeastern.edu}}

\begin{abstract}
We extend current models of the halo occupation distribution (HOD) to include a flexible, empirical framework for the forward modeling of the intrinsic alignment (IA) of galaxies. A primary goal of this work is to produce mock galaxy catalogs for the purpose of validating existing models and methods for the mitigation of IA in weak lensing measurements. This technique can also be used to produce new, simulation-based predictions for IA and galaxy clustering. Our model is probabilistically formulated, and rests upon the assumption that the orientations of galaxies exhibit a correlation with their host dark matter (sub)halo orientation or with their position within the halo.  We examine the necessary components and phenomenology of such a model by considering the alignments between (sub)halos in a cosmological dark matter only simulation.  We then validate this model for a realistic galaxy population in a set of simulations in the IllustrisTNG suite. We create an HOD mock with TNG-like correlations using our method, constraining the associated IA model parameters, with the $\chi^2_{\rm dof}$ between our model's correlations and those of Illustris matching as closely as 1.4 and 1.1 for orientation--position and orientation--orientation correlation functions, respectively. By modeling the misalignment between galaxies and their host halo, we show that the 3-dimensional two-point position and orientation correlation functions of simulated (sub)halos and galaxies can be accurately reproduced from quasi-linear scales down to $0.1~h^{-1}{\rm Mpc}$. We also find evidence for environmental influence on IA within a halo. Our publicly-available software provides a key component enabling efficient determination of Bayesian posteriors on IA model parameters using observational measurements of galaxy-orientation correlation functions in the highly nonlinear regime.
\end{abstract}

\keywords{%
Cosmology, Weak Gravitational Lensing, Intrinsic Alignments
}

\maketitle

\section{Introduction}


Weak lensing is an important probe for cosmological galaxy imaging surveys, and will be of major interest for the Rubin Observatory's Legacy Survey of Space and Time \citep[LSST,][]{Ivezic2019} as well as other Stage IV surveys such as {\it Euclid} \citep{Euclid2022} and {\it Roman} \citep{akeson2019wide}. These new surveys will be much more powerful than those in the past, and will require correspondingly more precise analyses. The intrinsic alignments (IA) exhibited by galaxies as a result of their alignment with large-scale structure and other nearby galaxies can contaminate cosmic shear measurements, driving the need for accurate IA modeling \citep{heymans_2006, Blazek_2011, Joachimi_2013, Krause_2016}.

IA has typically been modeled using analytic approaches, usually motivated through cosmological perturbation theory \citep[e.g.][]{Hirata_2004,Bridle_2007,Blazek_2015,Blazek_2019, vlah_2020, vlah_2021, maion_2023, bakx_2023, chen_2023}. However, these approaches are unable to describe alignments in the fully nonlinear regime. To address this issue,
in recent years, interest in halo-based models of the intrinsic alignment of galaxies has taken off, with \citet{Fortuna_2020} updating previous work by \citet{Schneider:2010cg} to provide a robust theoretical description of intrinsic alignment within dark matter halos. At the same time, efforts have emerged to infuse dark-matter-only N-body simulations with mock galaxies which have correlated alignments.
These methods include halo-based modeling \citep[e.g.][]{Joachimi_2013,Hoffmann_2022}, tidal field descriptions \citep[e.g.][]{Harnois-Deraps_2022}, and machine learning approaches \citep[e.g.][]{Jagvaral_2022}.
These simulation capabilities are important for validating our approaches to mitigate IA effects in weak lensing surveys, providing realistically complex mock data on which to test our analysis pipelines. These simulations can also be used to directly model IA, including in situations where analytic modeling is not feasible, such as higher-order or map-based analyses. Finally, the resulting mock galaxy catalogs can also help us in understanding and testing galaxy formation and evolution models which could impact the large-scale alignment of galaxies; either directly through comparison and calibration to hydro sims, or through interpreting the resulting behavior of different galaxy formation or evolution models in terms of the HOD IA parameterization. This mock generation process provides the connection between small scale galaxy formation physics either as calculated by a hydro sim or through some other analytic method and IA observables on a wide range of scales. In other words, the parameters of this HOD model are informed by the small-scale galaxy formation physics and allow us to make predictions, including statistical uncertainty, for a wide range of IA observables \citep{Hoffmann_2022, samuroff_etal23_des_ia_lum_color}.

In this work, we provide a new, flexible, and efficient halo-based method for including IA in mock galaxy catalogs. This method is built on the \href{https://github.com/astropy/halotools}{\halotools}\footnote{https://github.com/astropy/halotools} package for generating mock galaxy catalogs \citep{Hearin_2017} and can be applied within larger simulation and analysis frameworks to produce mock galaxy catalogs or for IA modeling \citep{van_alfen_joss}.

Models of the Halo Occupation Distribution (HOD) typically have components to capture the number of galaxies within halos (occupation), as well as the distribution of their positions and velocities within halos (phase space). In this work, we add two new components to this framework: one component to model the statistical distribution of galaxy orientations within their halos, and a second component, the IA strength model, that characterizes the probabilistic alignment between the galaxy orientation and some reference vector associated with the parent halo. These new ingredients to HOD modeling are quite flexible and allow for multivariate dependence on galaxy properties.

This paper is organised as follows: Section \ref{sec:two_point_clustering} presents background on two-point clustering, giving the forms of the correlation functions used to analyze our models. Section \ref{sec:simulations} describes the cosmologies of the simulations used in this work and discusses how we determine our covariance. In Section \ref{sec:model_misalignments} we introduce the Dimroth-Watson distribution that we use to characterize the distribution of galaxy orientations within their halos, and we describe the probabilistic models of central and satellite galaxy alignments that will be used throughout the rest of the paper. In Section \ref{sec:galaxy_halo_misalignment}, we examine the impact of altering central and satellite alignment strengths. This provides a baseline for understanding the simplest effect our alignment strengths have on our model. In Section \ref{sec:no_subhalos}, we use high-resolution N-body simulations to investigate the accuracy with which subhalo-orientation correlation functions can be captured by our HOD-type models using radial alignments, as subhalo orientations are often not available. We use the IllustrisTNG hydrodynamical simulation to test the flexibility of our model of alignment strength in Section \ref{sec:illustris}. Section \ref{sec:conclusions} summarizes our results and future work.

\section{Two-Point Correlation Functions and the Halo Occupation Distribution}
\label{sec:two_point_clustering}

In this section, we introduce the basic equations of the analytical model of halo occupation statistics. Historically, these analytic expressions have provided the principal framework underlying theoretical predictions of two-point galaxy clustering in the nonlinear regime \citep[e.g.,][]{berlind_etal02_hod,yang_etal03_clf}. In the present work, the theoretical prediction pipeline we introduce is purely simulation-based, and so our Monte Carlo methods for computing the integrals in this section are quite distinct from the algorithms implemented in other widely-used halo model libraries such as the Core Cosmology Library \citep[CCL,][]{chisari_etal19_ccl} or HMCode \citep{mead_etal15_hmcode}. We defer a comparative discussion of these methodologies to Section \ref{sec:conclusions}, and we include the analytical expressions here to elucidate the connection between simulation-based and analytical pipelines. Section \ref{sec:halo_theory} is purely pedagogical. None of the equations listed there are used in our work and are presented only to give context and background. The limitations of these equations and their necessary assumptions also highlight the importance of simulation-based methods. While these analytic equations rely on assumptions that only hold in certain regimes, a simulation-based approach needs no such assumptions. The following subsection, Section \ref{sec:estimators}, discusses the estimators used by our code to measure the correlations in our simulations.

\subsection{Theory: The Halo Model and correlation functions}
\label{sec:halo_theory}

In this paper, we consider the halo model \citep{cooray_2002, asgari_2023}, which provides a flexible approach for us to use in our creation of mock galaxy catalogs. Here we overview the theoretical equations for the two-point correlation functions. The functions are given to provide a conceptual background, but our analysis does not use these formulae to calculate correlation functions. Instead, we measure the correlations from mock data using the estimators described in Section \ref{sec:estimators}.

Below, we see the theoretical formulation of the position-position correlations.

\beq
\tpcf{gg}(r) = \tpcftwo{gg}{1h}(r) + \tpcftwo{gg}{2h}(r)
\eeq

\beq
\tpcftwo{gg}{1h}(r) = \tpcftwo{cs}{1h}(r) + \tpcftwo{ss}{1h}(r) 
\eeq

\beq
\tpcftwo{gg}{2h}(r) =  \tpcftwo{cc}{2h}(r) + \tpcftwo{cs}{2h}(r) + \tpcftwo{ss}{2h}(r) 
\eeq

\begin{align}
\label{eq:tpcfgg2h}
\begin{split}
\tpcftwo{gg}{2h}(r) =   \frac{1}{\bar{n}_{\rm g}^{2}}\int\dd M_1 \int\dd M_2 & \frac{\rm dn}{\dd M_1}\frac{\rm dn}{\dd M_2} \tpcf{hh}(r\vert M_1, M_2) \\& \times  \meantwo{N_{\rm g}}{M_1} \meantwo{N_{\rm g}}{M_2},
\end{split}
\end{align}

\noindent where $\frac{\rm dn}{\dd M}$ is the {\it halo mass function} (the differential distribution of number density of halos $n$ as a function of halo mass $M$), $\meantwo{N_{\rm g}}{M}$ is the expectation value of the number of galaxies in a halo of mass $M$, and $\bar{n}_{\rm g}$ is the overall number density of the population of galaxies being considered, and we have ignored the central and satellite fractions, which are often absorbed into the definition of the correlation function. In the above equations, subscripts indicate the two populations being correlated. The subscript ``g" refers to all galaxies, ``h" refers to halo, ``c" refers to central galaxy (the most massive galaxy at/near the center of the host dark matter halo), and ``s" refers to satellite galaxy (a smaller galaxy in a host dark matter halo besides the central galaxy). The superscripts ``1h" and ``2h" denote the scale contributing to the correlations. Namely, ``1h" refers to the 1-halo regime (galaxies within the same halo), ``2h" refers to the 2-halo regime (galaxies residing in separate halos), and no superscript indicates correlations from all scales. In this paper, we use 1 Mpc/$h$ as the approximate delineation between the 1-halo and 2-halo regimes.

The assumptions underlying Eq.~\eqref{eq:tpcfgg2h} are:
\bit
\item $r\gg R_{\rm vir}^{\rm max}$
\item On very large scales, $\tpcf{hh}(r)$ varies slowly across the length scale of halos
\eit
where $R_{\rm vir}^{\rm max}$ refers to the largest virial radius of all halos being considered for the 2-halo contribution to the correlation. These assumptions together allow equation \eqref{eq:tpcfgg2h} to treat halos as a single point with a given number density of galaxies concentrated at that point. Clearly, while these assumptions work at large scales, they break down as separations approach the 1-halo regime. At such scales, equation \eqref{eq:tpcfgg2h} no longer accurately represents the correlation function. The breakdown of this assumption is an example of where simulation-based methods are useful.

\begin{align}
\label{eq:tpcfgg1h}
\begin{split}
\tpcftwo{gg}{1h}(r) =  \frac{1}{\bar{n}_{\rm g}^{2}} \int\dd M \frac{\rm dn}{\dd M} \int\dd^{3}& x\lambda(\vec{x}\vert M)\lambda(\vec{x}+\vec{r}\vert M)\\ & \times  \meantwo{N_{\rm g}(N_{\rm g}-1)}{M}
\end{split}
\end{align}

In Eq.~\eqref{eq:tpcfgg1h}, $\lambda(\vec{x}\vert M)$ is the {\em galaxy profile}, i.e., the unit-normalized spatial distribution of galaxies within a halo of mass $M$. This expression includes both central--satellite and satellite--satellite pairs.

Above, we looked at the position-position correlation functions. Here, we overview the orientation-orientation and position-orientation correlation functions. The equations presented here are analogous to those described in more detail in \citet{Fortuna_2020}. Following typical naming conventions \citep[e.g.][]{lee_2008, tenneti_2015}, we use the subscripts ``e" and ``d" to refer to ellipticity and direction respectively, where direction refers to the separation between galaxies, which in some contexts we call ``position.'' As discussed below, we do not use full shape information in this work, so our use of ``ellipticity" refers in practice to orientation.

\begin{align}
\label{eq:tpcfee2h}
\begin{split}
\tpcftwo{ee}{2h}(r) =   \frac{1}{\bar{n}_{\rm g}^{2}}\int\dd M_1& \int\dd M_2  \frac{\rm dn}{\dd M_1}\frac{\rm dn}{\dd M_2} \tpcf{hh}(r\vert M_1, M_2) \\& \times \meantwo{N_{\rm g}}{M_1} \meantwo{N_{\rm g}}{M_2} \mean{\vert\evec{1}\cdot\evec{2}\vert^{2}}
\end{split}
\end{align}

In Eq.~\eqref{eq:tpcfee2h}, $\vec{e}$ is the three-dimensional ellipticity or orientation (i.e.\ the unit-normalized ellipticity) of the galaxy. 
Similarly, the 1-halo part of the ee correlation function and both parts of the ed correlation function are

\begin{align}
\label{eq:tpcfee1h}
\begin{split}
\tpcftwo{ee}{1h}(r) =  \frac{1}{\bar{n}_{\rm g}^{2}} \int\dd M& \frac{\rm dn}{\dd M} \int\dd^{3} x\lambda(\vec{x}\vert M)\lambda(\vec{x}+\vec{r}\vert M)\\ & \times \meantwo{N_{\rm g}(N_{\rm g}-1)}{M}   \mean{\vert\evec{1}\cdot\evec{2}\vert^{2}},
\end{split}
\end{align}

\begin{align}
\label{eq:tpcfed2h}
\begin{split}
\tpcftwo{ed}{2h}(r) =   &\frac{1}{\bar{n}_{\rm g}^{2}}\int\dd M_1 \int\dd M_2  \frac{\rm dn}{\dd M_1}\frac{\rm dn}{\dd M_2} \tpcf{hh}(r\vert M_1, M_2) \\& \times \meantwo{N_{\rm g}}{M_1} \meantwo{N_{\rm g}}{M_2} \left( \mean{\vert\evec{1}\cdot\hat{r}\vert^{2}} +  \mean{\vert\evec{2}\cdot\hat{r}\vert^{2}} \right),
\end{split}
\end{align}

and

\begin{align}
\label{eq:tpcfed1h}
\begin{split}
\tpcftwo{ed}{1h}(r) =  &\frac{1}{\bar{n}_{\rm g}^{2}} \int\dd M \frac{\rm dn}{\dd M} \int\dd^{3} x\lambda(\vec{x}\vert M)\lambda(\vec{x}+\vec{r}\vert M)\\ & \times \meantwo{N_{\rm g}(N_{\rm g}-1)}{M} \left(  \mean{\vert\evec{1}\cdot\hat{r}\vert^{2}} + \mean{\vert\evec{2}\cdot\hat{r}\vert^{2}} \right).
\end{split}
\end{align}

For $\tpcf{ee}(r),$ we can decompose the final term in the integrand into central and satellite contributions:

\begin{align}
\begin{split}
\mean{\vert\evec{1}\cdot\evec{2}\vert^{2}} &= 
f_{\rm cen, 1}f_{\rm cen, 2}\mean{\vert\evec{cen, 1}\cdot\evec{cen, 2}\vert^{2}} \\
& + f_{\rm sat, 1}f_{\rm sat, 2}\mean{\vert\evec{sat, 1}\cdot\evec{sat, 2}\vert^{2}} \\
&+ f_{\rm cen, 1}f_{\rm sat, 2}\mean{\vert\evec{cen, 1}\cdot\evec{sat, 2}\vert^{2}} \\
& + f_{\rm sat, 1}f_{\rm cen, 2}\mean{\vert\evec{sat, 1}\cdot\evec{cen, 2}\vert^{2}}  
\end{split}
\end{align}
where $f_{\rm cen, \it i}$ and $f_{\rm sat, \it i}$ are the fraction of galaxies in sample $i$ that are centrals and satellites, so that $f_{\rm cen, \it i} + f_{\rm sat, \it i}=1.$ 

For $\tpcf{ed}(r),$ this central-satellite decomposition takes the form:
\begin{align}
\begin{split}
 \mean{\vert\evec{i}\cdot\hat{r}\vert^{2}} =   f_{\rm cen, \it i}\mean{\vert\evec{cen, \it i}\cdot\hat{r}\vert^{2}} +  f_{\rm sat, \it i}\mean{\vert\evec{sat, \it i}\cdot\hat{r}\vert^{2}}
\end{split}
\end{align}

\noindent where $\vec{r} = r\cdot\hat{r}$ refers to the position vector between two galaxies.

\subsection{Two-point Correlation Functions Estimators}
\label{sec:estimators}

Below are the estimators for the position-position, position-orientation, and orientation-orientation correlation functions that we use to analyze our model. While earlier we showed the theoretical formulation, here we focus on the estimators we use on mock galaxy catalogs.
As stated earlier, to remain consistent with the names used in the literature, we have described these functions using ``ellipticity'' rather than ``orientation.'' However, while Eqs.~\eqref{eq:tpcfee2h}-\eqref{eq:tpcfed1h} can use full ellipticity, the rest of our paper and the estimators below only use the galaxy orientation. Our method, as presented here, does not include full galaxy shape information, which is the topic of ongoing work. We also note that the functions utilized here are convenient for analysing simulations where three-dimensional position and orientation information is available. However, for analysis of observed data, where only projected shape information is present, other correlation functions are typically used \cite[e.g.][]{Mandelbaum_2006, singh2023increasing}.

The galaxy-galaxy two-point correlation function is given by:
\begin{equation}
\label{eq:xi-estimator}
\xi_{gg}(r) = \frac{DD(r)}{RR(r)} - 1
\end{equation}
where $DD(r)$ is the number of galaxy pairs separated by $r$ and $RR(r)$ is the expected number of pairs for a random distribution. We use this estimator rather than the Landy-Szalay estimator \citep{landy_1993} even though it is somewhat sub-optimal in some cases \citep{sukhdeep_2017}. This estimator is faster, and the periodic nature of our box allows us to use analytic randoms, negating much of the sub-optimality.

The ellipticity-direction (ED) correlation function is defined as: 
\begin{equation}
\label{eq:omega-estimator}
\omega(r) = \langle |\hat{e}({\bf x}) \cdot \hat{r}|^2 \rangle -\frac{1}{3}.
\end{equation}

The ellipticity-ellipticity (EE) correlation function is defined as: 
\begin{equation}
\label{eq:eta-estimator}
\eta(r) = \langle |\hat{e}({\bf x}) \cdot \hat{e}({\bf x}+{\bf r})|^2 \rangle -\frac{1}{3},
\end{equation}

where ${\bf x}$ is the position vector of a given galaxy, ${\bf r}$ is the separation vector between two galaxies, and $\hat{r}$ is the unit vector of the separation vector ${\bf r}$. In both of these equations, subtracting the factor of $1/3$ accounts for the fact that integrating the dot product of the two vectors (i.e. integrating the $\cos^2{\theta}$ between them) over a sphere results in a factor of $1/3$ when the vectors are randomly distributed (i.e. there is no correlation). These equations are examined in more depth in \citet{chisari_2016} with discussion about the factor of $1/3$ in \citet{Hopkins_2005}.

Beyond this section, the symbols for equations \eqref{eq:xi-estimator}-\eqref{eq:eta-estimator} will be used. Namely we use $\xi$ for the position--position correlation function, $\omega$ for the position--orientation correlation function, and $\eta$ for the orientation--orientation correlation function. We do this to separate the rest of our work from the pedagogical discussion here as well as to connect the measurements discussed later with the estimators used to obtain them.

\section{Simulations and Covariance}
\label{sec:simulations}

We construct our HOD models using {\tt halotools}, a library for simulation-based predictions of the galaxy--halo connection. Catalogs of dark matter halos are the root data product underlying {\tt halotools}-based predictions, and throughout the paper we make use of catalogs provided by the library, although we note that {\tt halotools} provides full support for user-supplied simulation data. Upon ingesting a catalog of simulated halos, {\tt halotools} populates each halo with a synthetic galaxy population according to an occupation model, distributing galaxies within each halo according to an assumed phase space model. Orientations are bestowed upon each synthetic galaxy according to the alignment models outlined in Section \ref{sec:model_misalignments}.

Throughout the paper, we will distinguish between a {\em host halo}, which is a distinct gravitationally self-bound collection of simulated dark matter particles, and a {\em subhalo}, a gravitationally self-bound object that is located inside the boundary of a larger host halo. We will use the term {\em (sub)halo} whenever referring to a collection of host halos and subhalos together and/or when referring to an individual object that may be either a subhalo or a host halo.

We have previously defined central galaxies as the most massive galaxy in a dark matter halo, located at or near the center of the halo. Satellite galaxies are all galaxies in a halo other than the central galaxy. The ratio of central galaxies to satellite galaxies depends on the occupation parameters of the HOD model. As a rule of thumb, the galaxy tables produced with {\halotools} using occupation parameters typical of those this work tend to contain roughly three central galaxies to every one satellite galaxy.

\subsection{Simulations}

\begin{table*}[!ht]
    \centering
    \begin{tabular}{cccccccc}
        \hline
        Simulation & Particle Mass & $\Omega_{m,0}$ & $\sigma_8$ & $n_s$ & $h$ & $L_{\rm box}$ & $z$\\
         & ($h^{-1} M_{\odot}$) & & & & & ($h^{-1}$ Mpc) & \\
        \hline \hline
        Bolshoi-Planck & $\sim10^8$ & 0.30711 & 0.82 & 0.96 & 0.70 & 250 & 0\\
        (BolPlanck) & & & & & & & \\
        \hline
        Small MultiDark Planck & $\sim10^8$ & 0.307115 & 0.8228 & 0.96 & 0.6777 & 400 & 0\\
        (SMDPL) & & & & & & & \\
        \hline
        IllustrisTNG300 & $\sim10^{10}$ & 0.3089 & 0.8159 & 0.9667 & 0.6774 & 205 & 0\\
        \hline
    \end{tabular}
    \caption{Cosmology and other parameters for the three simulations described in Section \ref{sec:simulations}. Additional information about each simulation can be found in said section.}
    \label{tab:simulations}
\end{table*}

In this paper, unless otherwise stated, whenever we create a catalog using {\tt halotools}, we do so using the available halo catalogs from the Bolshoi-Planck\footnote{\href{https://doi.org/10.17876/cosmosim/bolshoip}{https://doi.org/10.17876/cosmosim/bolshoip}} (BolPlanck) simulation output at $z=0$, the cosmology and simulation parameters of which are given in Table \ref{tab:simulations} \citep{Klypin_2011}.

While most of the paper uses BolPlanck, Figures \ref{fig:degrading_central_alignments} and \ref{fig:degrading_satellite_alignments} use the Small MultiDark Planck (SMDPL) simulation with cosmology described in Table \ref{tab:simulations}. The simulation is evolved by solving for gravitational interactions only using the {\tt L-GADGET-2} code, a version of the publicly available cosmological code {\tt GADGET-2} \citep{2005MNRAS.364.1105S} with a force resolution of $1.5~h^{-1} {\rm kpc}$.  This simulation belongs to the series of MultiDark simulations with Planck cosmology.  More details for this simulation are described in \citet{2016MNRAS.457.4340K}.

For both BolPlanck and SMDPL, (sub)halos are found using the phase-space halo finder {\tt ROCKSTAR} \citep{2013ApJ...762..109B}, which uses adaptive, hierarchical refinement of friends-of-friends groups in six phase-space dimensions and one time dimension, and tracked over time using the {\tt Consistent Trees} algorithm \citep{2013ApJ...763...18B}. As demonstrated in \citet{2011MNRAS.415.2293K, 2013MNRAS.435.1618K}, this results in a very robust tracking of (sub)halos \citep[also see][]{2016MNRAS.458.2848J}.

The essential orientation properties used in this paper (the x, y, and z values of the unit vector for the major axis of the ellipsoid) are also generated with the {\tt ROCKSTAR} halo finder, shown in Appendix B of \citet{rodriguez_puebla_2016}. For a more detailed look at shape finding, see Appendix \ref{appendix:shapes}.

ROCKSTAR determines virial spherical overdensity (SO hereafter) volumes for each halo, centered on a local density peak (possible for even very ellipsoidal halos), such that the average density inside the sphere is $\bar{\rho}_h(z) = \Delta_{\rm vir}(z)\rho_{\rm m}(z)$. Here $\rho_{\rm m}(z) = \Omega_{\rm m}(z)\rho_{\rm crit}(z)$, where $\rho_{\rm crit}(z) = 3H(z)2/8\pi G$ is the critical energy density of the Universe, and $\Delta_{\rm vir}(z)$ is given by a fitting function \citep{1998ApJ...495...80B}:
\begin{equation}
   \Delta_{\rm vir}(z)= \left[ 18\pi^2-82\Omega_{\Lambda}(z)-39\Omega_{\Lambda}^2(z)\right]\Omega_{\rm m}^{-1}
\end{equation}
For the Planck cosmology, $\Delta_{\rm vir}(z)\simeq 360$. The radius of each such sphere defines the virial radius $R_{\rm vir}$ of the halo, which is related to the mass of the halo via $M_{\rm vir} = (4/3)\pi R_{\rm vir}^3 \bar{\rho}_{h}$. Additionally, subhalos in this catalog are distinct, self-bound structures whose centers are found within the virial radius of a more massive host halo.

In Section \ref{sec:illustris}, we compare results from our HOD models (built using the BolPlanck catalogs described above) to the Next Generation Illustris Simulations\footnote{\href{http://www.tng-project.org}{http://www.tng-project.org}} (IllustrisTNG), a suite of hydrodynamical simulations of galaxy formation in cosmological volumes \citep{nelson_2017, pillepich_2017, naiman_2017, springel_2017, marinacci_2017}. Illustris resolves galaxy shapes, therefore the IA seen in Illustris is modeled directly by hydrodynamic interactions. The IllustrisTNG runs used in this work are two uniform mass resolution cosmological volume simulations with side lengths 205 $h^{-1}{\rm Mpc}$, one ``full physics" run, TNG300, including all of the complex physics of galaxy formation, and a gravity-only counterpart, TNG300-Dark. The initial conditions of the simulations were set at z = 127 using the Zeldovich approximation. The adopted cosmological parameters are given in Table \ref{tab:simulations}.

It is unclear exactly to what extent the cosmologies of the underlying halo catalogs shown in Table \ref{tab:simulations} will affect resulting galaxy tables. However, the methods described here are highly tunable and can produce a wide range of galaxy populations and alignments from a single halo catalog as shown in Section \ref{sec:illustris}.

\subsection{Covariance}
\label{sec:covariance}

Building our HOD model for IA is inherently stochastic, as both the selection of halos being populated and the alignment of galaxies (discussed below) are drawn from distributions. When fitting this model to ``data'' (which in this work include both measurements of the simulated halos being populated as well as galaxies in another simulation such as IllustrisTNG), we must account for the fact that both the data and model have stochasticity. When fitting to the halo measurements (Section \ref{sec:no_subhalos}), we find that calculating the covariance using multiple realizations of the galaxy alignments, without changing the halo occupation, is sufficient to allow a test of the model performance while avoiding double counting the variance from the occupation step. When fitting to IllustrisTNG (Section \ref{sec:illustris}), we calculate the covariance from IllustrisTNG and average over multiple model realizations to remove model variance. Further details concerning our covariance are given in Appendix \ref{appendix:covariance}.

\section{Modeling Alignments}
\label{sec:model_misalignments}

In this paper, we make the assumption that both galaxies and (sub)halos can be modeled as triaxial homologous ellipsoids (see Appendix \ref{appendix:shapes} for details).  The orientation of galaxies/halos can then be entirely described by specifying the vectors associated with the minor, intermediate, and semi-major axes of ellipsoids.  Here we focus on modeling the orientation of a single axis of the ellipsoid describing a galaxy within its host halo, but we note that the following framework could be extended to model a three-dimensional galaxy orientation and shape, albeit with increased complexity.  

\begin{figure}[!ht]
    \centering
    \includegraphics{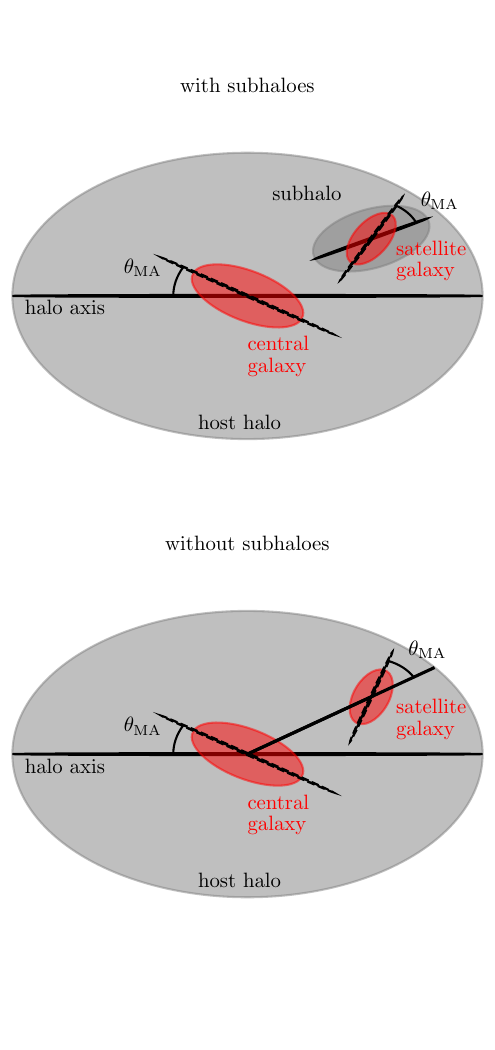}
    \caption{Cartoon of alignment model when subhalos with accurate shapes are available (top) and without subhalos (bottom). In both cases, the misalignment angle is determined with respect to some reference axis, which differs for satellite galaxies depending on the model used, either using subhalo alignment (top), or radial alignment (bottom).}
    \label{fig:model_cartoon}
\end{figure}

Our strategy will be to statistically model the misalignment angle between a vector specifying the orientation of each galaxy's (sub)halo, and a vector specifying the orientation of the galaxy, as shown in Figure \ref{fig:model_cartoon}.  Mathematically, if we assume that the vector specifying the orientation of each galaxy's host (sub)halo is the unit vector, $\hat{z}$, we can model the orientation vector of the galaxy by specifying the angular coordinates on the unit-sphere.

For this purpose, we utilize the Dimroth-Watson distribution, which specifies the spherical polar coordinates of a unit vector. We chose the Dimroth-Watson distribution as it provides a maximum entropy distribution on a sphere while accounting for the 180 degree symmetry we see in our galaxy orientations \citep{Watson_1965}. These criteria are difficult to meet with other distributions, but the Dimroth-Watson distribution naturally fits the role for directional, axially symmetric systems on a sphere \citep{sra_2013}. The probability distribution for the polar angle, $\theta$, and the azimuthal angle, $\phi$, are given by
\begin{equation}
P(\theta,\phi) = \frac{B(\kappa)}{2\pi}e^{-\kappa\cos^2(\theta)}\sin(\theta)\mathrm{d}\theta\mathrm{d}\phi
\label{eq:watson}
\end{equation}
where the normalization factor is given by:
\begin{equation}
B(\kappa) = \frac{1}{2}\int_0^1 e^{-\kappa t^2}\mathrm{d}t
\end{equation}
Note that the azimuthal angle, $\phi$, is modeled as a uniform distribution, so that the angle $\theta$ is then by definition the misalignment angle between galaxy and host (sub)halo orientation, $\theta_{\rm MA}$.  The degree to which galaxies and (sub)halos align is then controlled by the $\kappa$ parameter.   

\begin{figure}[!ht]
    \includegraphics[width=\columnwidth]{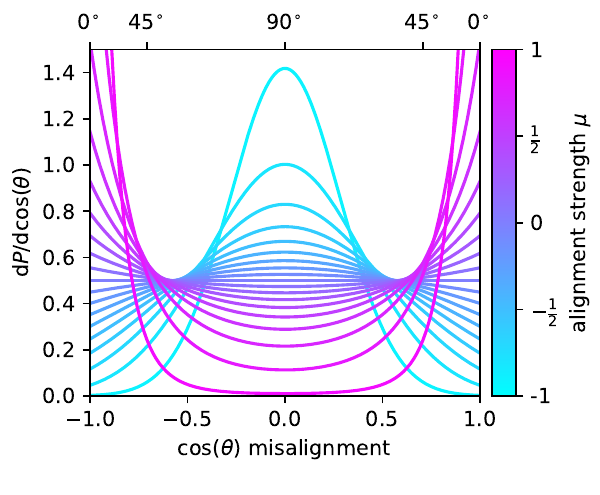}
    \caption{Dimroth-Watson distribution model, equation~\eqref{eq:watson}, for the distribution of galaxy--halo orientation misalignment angles, $\theta$.  Different color lines show distributions with different alignment strengths, $\mu$, given by equation \eqref{eq:alignment_strength}. We see that as $\mu$ approaches 1, the Dimroth-Watson distribution approaches delta functions at $\cos{\theta} = 1$ and $-1$ (perfect alignment). Meanwhile, as $\mu$ approaches $-1$, the distribution approaches a delta function at $\cos{\theta} = 0$ (perfect anti-alignment). Notably, the value of $\mu$ does not deterministically assign misalignment angle (except in the cases of perfect alignment and perfect anti-alignment), but rather it determines the shape of the distribution from which misalignment angles will be drawn.}
    \label{fig:dimroth_watson}
\end{figure}

It is convenient to re-parameterize the strength of alignment $\kappa$ by:
\begin{equation}
\label{eq:alignment_strength}
\mu = \frac{-2 \tan^{-1}(\kappa)}{\pi}
\end{equation}
such that $\mu=1, 0, -1$ corresponds to perfect alignment, random alignment, and perfect anti-alignment, respectively.  As illustrated in Figure \ref{fig:dimroth_watson}, a positive value of $\mu$ corresponds to a preferential parallel alignment between a galaxy and its host (sub)halo.  A negative value of $\mu$ corresponds to a preferential perpendicular alignment between a galaxy and its host halo.   

The distribution of misalignment angles is symmetric about $\cos(\theta_{\rm MA})=0$, in accord with the symmetries we have assumed of the shapes of galaxies and halos. There is thus no difference in the morphological orientations between a galaxy rotated by $180^{\circ}$ and its original orientation.  In principle, for disk galaxies whose orientation is determined by angular momentum, it may be interesting to distinguish between parallel and anti-parallel angular momentum vectors. Although the difference should not impact the observable galaxy ellipticities, correlations between the angular momentum direction and the density field can allow us to probe tidal torque theory (e.g.\ \citealp{lee2019}) -- we leave exploration of this and related effects to future work.

\subsection{Galaxy Alignment Models}

For all investigations, we align the major axes of central galaxies with the major axes of their host halos.

For satellite galaxies, we explore two models in this paper for determining their orientations:
\begin{enumerate}
    \item satellites are oriented relative to that of their host dark matter subhalo, called {\em subhalo alignment} (as seen in the upper image of Figure \ref{fig:model_cartoon});
    \item satellites are oriented relative to the host halo centric radial vector, called {\em radial alignment} (illustrated in the lower image of Figure \ref{fig:model_cartoon}). In this case, there are two different alignment strengths explored.
    \begin{enumerate}
        \item Constant alignment strength, where the alignment strength is the same for all satellite galaxies.
        \item Distance-dependent alignment strength, where the alignment strength of a satellite depends on the distance to its central galaxy.
    \end{enumerate}
\end{enumerate}

In the case of subhalo alignment, the model needs information about subhalo positions and orientations, while the radial alignment model may be used with or without subhalo positions and needs no information about subhalo orientations (though we typically use subhalo positions unless otherwise stated).

\subsection{Building an HOD Mock}
To build an HOD-based mock catalog, we use the HODModelFactory in \halotools. Central and satellite galaxies have separately-defined occupation models that determine the number of galaxies of each population present in a given halo. Additionally, each population has its own model for the intra-halo phase space distribution. Throughout the paper, central galaxies are placed at the center of the host halo with the same peculiar velocity as the halo. For satellite galaxies, we consider two different classes of phase space models. The first is defined by the NFWPhaseSpace class: the spatial distribution of satellites is spherically symmetric and has a radial density profile given by the NFW distribution \citep{navarro_frenk_white_1997_nfw_profile}; satellite velocities are isotropic and exhibit a velocity dispersion profile determined by the Jeans equation. The second satellite model we consider is given by the SubhaloPhaseSpace class, in which the position and velocity of each satellite is defined by a subhalo that has been randomly selected from the simulated halo, preferentially selecting the most massive subhalos first; when $N_{\rm sat}$ is smaller than the number of subhalos in the host halo, the lowest-mass subhalos are left unoccupied, and when $N_{\rm sat}$ exceeds the number of available subhalos, the remaining satellites are distributed according to the NFWPhaseSpace class. The SubhaloPhaseSpace class provides a way for the synthetic satellite population to inherit the complexity of the intra-halo distributions of simulated subhalos, while at the same time retaining the flexibility of the HOD to occupy host halos with a variable number of satellites.

\section{Impact of Galaxy--Halo Misalignment on Orientation Correlation Functions}
\label{sec:galaxy_halo_misalignment}

In this section, we use our halo model of IA to provide a pedagogical investigation of how galaxy--halo misalignment manifests in orientation correlation functions. We will separately study effects arising from central and satellite galaxies, and we will consider the full range of spatial scales, spanning both 1-halo and 2-halo regimes. Our general strategy for this investigation is to begin by orienting galaxies to be perfectly aligned with the reference vector of the (sub)halo; we then leverage the flexibility of our probabilistic model by programmatically weakening the alignment strength, and studying how the orientation correlation function changes as a result.

\subsection{Central and Satellite Galaxy Misalignment}
\label{sec:cen_sat_misalingments}
\begin{figure*}
    \centering
    \includegraphics[height=\columnwidth]{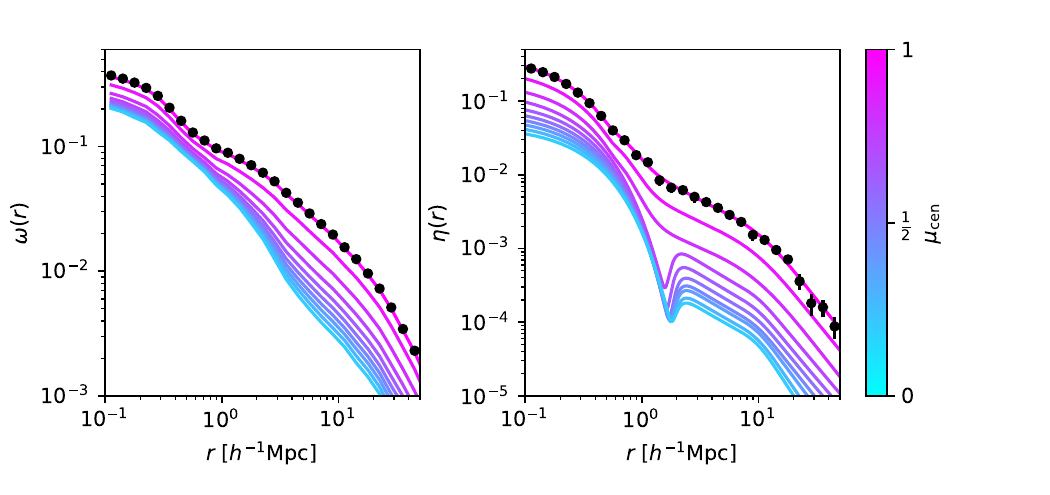}
    \caption{The effect of degrading the central galaxy-host halo alignment strength on the $\omega$ (left) and $\eta$ (right) orientation correlation functions.  The black points with error bars (which are often smaller than the points) are measurements made directly on (sub)halos.  The colored lines are for model galaxies with varying levels of alignment strength between central galaxies and their host halo, from random alignments (light blue) to perfect alignments (pink).  In each model, satellite galaxies take on the same orientation as their subhalo.  On the right, a fitting function is used to smooth the results, letting us focus on the overall trend from lowering the central alignments which tends to be noisy for $\omega$ correlations. We also see a downward bump around the 1-halo to 2-halo scale transition in the right panel. This suggests to us an effect related to the satellite--satellite correlations, with the sudden upturn going into the1-halo regime being related to satellite pairs within a halo becoming more significant. This effect also only becomes apparent once the central alignment strength has become weak enough that the satellite alignments become more important.}
    \label{fig:degrading_central_alignments}
\end{figure*}

We begin by examining the effect of varying the central galaxy -- host halo alignment strength on the orientation correlation functions.  In Figure \ref{fig:degrading_central_alignments}, we show the position--orientation ($\omega$) on the left and orientation--orientation ($\eta$) correlation functions on the right. In both plots, we show subhalos and host halos with a peak mass greater than $10^{12}~h^{-1} M_{\odot}$ as black points with error bars.  Errors on these measurements are estimated using jackknife re-sampling of the simulation box, by splitting into $4^3$ equal volume sub-samples.

Also shown in Figure \ref{fig:degrading_central_alignments} are color-coded lines, showing $\omega^{\rm g}_{\rm gg}$ and $\eta^{\rm g}_{\rm gg}$ (position-orientation and orientation-orientation correlations for all galaxies) for model galaxies in the same sample of (sub)halos with varying levels of central-host halo major axis alignments. Here we use the superscripts ``g" and ``h" to specify the correlation is being done on the galaxies or halos respectively (e.g. $\omega^{\rm h}_{\rm cs}$ refers to the position--orientation correlation between the positions of halos hosting central galaxies and the orientations of subhalos hosting satellite galaxies). The satellite--subhalo alignment strength is kept fixed at full strength, $\mu_{\rm sat}=1$.  Because there is still significant uncertainty in the $\eta$ measurement in a $400~{\rm Mpc}$ box, we use a fitting function to smooth the results (see Appendix \ref{appendix:fitting_functions} for details).  For maximum alignment, $\mu_{\rm cen}=1$, the correlation functions reproduce the (sub)halo measurement by construction.

\begin{figure*}
    \includegraphics[height=\columnwidth]{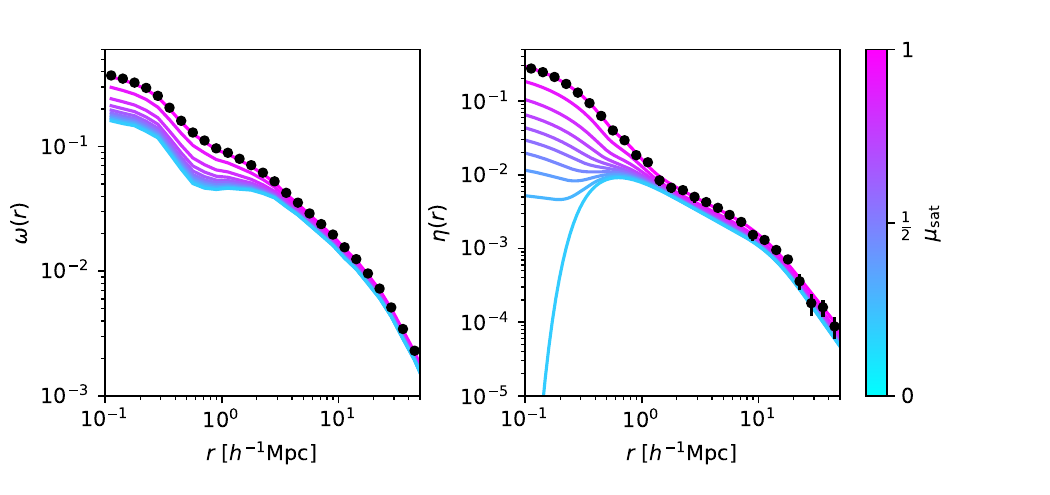}
    \caption{The effect of degrading the satellite galaxy-subhalo alignment strength on the $\omega$ (left) and $\eta$ (right) orientation correlation functions.  The black points with error bars are measurements made directly on (sub)halos.  The colored lines are for model galaxies with varying levels of alignment strength between satellite galaxies and their subhalo, from random alignments (light blue) to perfect alignments (pink).  In each model, central galaxies take on the same orientation as their host halo.  On the right, a fitting function is used to smooth the results.}
    \label{fig:degrading_satellite_alignments}
\end{figure*}

Next we show the effect of varying the satellite--subhalo alignment strength on the orientation correlation functions.  Similar to previous section, in Figure \ref{fig:degrading_satellite_alignments} we show the $\omega$ on the left and $\eta$ on the right, with correlation functions for subhalos and host halos given as black points with error bars. In Figure \ref{fig:degrading_satellite_alignments}, the color-coded lines correspond to varying levels of satellite--subhalo alignments, while the central--halo alignment strength is kept fixed at full strength, $\mu_{\rm cen}=1$. Again, when $\mu_{\rm sat}=1$, the (sub)halo orientation correlation functions are reproduced by construction.

Broadly speaking, we can see that central galaxy alignments produce a much larger effect on the correlation functions relative to satellites, especially at larger scales. The satellite correlations in Figure \ref{fig:degrading_satellite_alignments} show noticeable effects at smaller scales, but become much less important on scales larger than $\sim1$ Mpc/$h$. Meanwhile, Figure \ref{fig:degrading_central_alignments} shows that changing the alignment strength of centrals has a noticeable effect at all scales.

\section{Modeling Satellite Orientations without Subhalo Orientations}
\label{sec:no_subhalos}

Our analyses in the previous sections have relied upon alignments in which the orientation of a synthetic satellite was variably-correlated with the major axis of its parent subhalo. But relying on subhalo catalogs as the core data product of the model introduces a significant source of systematic error arising from uncertainty in subhalo-finding and numerical/resolution issues on small scales \citep[e.g.,][]{vdb_ogiya_etal2018_disruption,campbell_etal18_sham_crisis}. The resolution requirements on subhalo finding are steep, especially for predictions involving subhalo internal structure \citep[e.g.,][]{rodriguez_puebla_2016, benson_2017_noisefree_spin, mansfield_avestruz_2021}, and so making theoretical predictions with models that require resolved subhalo orientations leads to costly demands on the Gpc-sized simulations that are required to analyze cosmological surveys. The positions of the subhalos themselves are fairly well-resolved and do not introduce notable uncertainty.

These considerations motivate an alternative formulation of satellite orientations with various levels of subhalo information. We start by looking at radial alignments for cases where no subhalo orientation is present, while still using subhalo positions to place satellites. From there, we move on to examine the ability of this simple model to match halo correlations in the 1-halo and 2-halo regimes, justifying our choice to fit our parameters to the 1-halo regime. Following that discussion, we examine the effect of varying levels of satellite anisotropy, including a fully isotropic model that shows the effect of building an HOD with no subhalo information at all. We then conclude this section with a discussion on the goodness of fit of the best radial alignment models as well as by introducing an alternate method of determining a radial alignment strength in the case where subhalo orientations are present.

In this section we carry out a targeted study of the flexibility of IA models of satellites that do not rely upon subhalo orientations.  Our primary assumption will be that the orientation of a satellite galaxy tends to align radially towards the center of the host halo.  We examine this assumption in the simplest alignment model in \S \ref{sec:radial_alignment}, and in subsequent subsections we proceed to incorporate additional complexity and study the impact on orientation correlation functions. In order to assess the flexibility required by an HOD-type model of satellite orientations, we begin by focusing on recovering the orientation correlation functions exhibited by (sub)halos at $z=0$ in the gravity-only Bolshoi-Planck simulation. Note that we do not claim that satellite galaxies always perfectly align with their respective subhalos, but we use this as a test of how much subhalo orientation information can be captured when such information is not available. In this way, we probe the full flexibility provided by our model. Following this preliminary investigation, in Section \ref{sec:illustris}, we will proceed to apply our IA model to full-physics hydrodynamical simulations (described in Section \ref{sec:simulations}).  

\subsection{Radial Alignment Modeling}
\label{sec:radial_alignment}

Constant radial misalignment uses a single, constant alignment strength for all galaxies, giving them all the same Dimroth-Watson distribution for their radial alignments. A distance-dependent alignment strength changes the shape of the Dimroth-Watson distribution for each satellite galaxy depending on its distance from its central galaxy. We choose a form of this relationship given by a power law -- equation \eqref{eq:galaxy_alignment_strength}. Because $\mu \in [-1,1]$, this equation is limited to that range as well, assigning -1 to all $\mu < -1$, and 1 to all $\mu > 1$.

\begin{equation}
    \mu = a\,(r/r_{\textrm{vir}})^{\gamma}
    \label{eq:galaxy_alignment_strength}
\end{equation}

Note that equation \eqref{eq:galaxy_alignment_strength} is not the only possible model for radial alignment, but it is one possible simple way to model distance-dependent radial alignments. In particular, this model requires that all satellite galaxies have either preferential alignment, or preferential anti-alignment (or random in the case of $a = 0$). We have chosen this particular model for its simplicity, but the same concept of assigning individual alignment strengths based on such a model is general.

Since $\mu$ refers to how strongly galaxies tend to align with respect to a given reference vector, the $\mu$ values in this section refer to how strongly satellite galaxies align with the radial vector between the satellite and the center of its host halo. In previous sections, we used (sub)halo major axes as reference vectors, so matching the halo correlation in that context would mean perfectly aligning with the halo major axes. Here, however, matching the halo correlations is less straightforward. We would only expect a value of $\mu=1$ in this case if subhalo shapes are perfectly aligned with their radial vectors.

In any given mock galaxy catalog that we generate, there are roughly three times as many central galaxies as there are satellite galaxies (i.e., the satellite fraction is roughly 25\% for the galaxy populations we consider), so looking at the correlation functions between all galaxies dilutes the effect of the satellites, and is heavily influenced by the central correlations. In order to compare the effects of these satellite alignments strengths, we look at $\omega^{\rm g}_{\rm cs}$ (the correlation functions for the central galaxy position and satellite galaxy orientation) rather than the full $\omega$ correlation function for all galaxies to all galaxies. In this way, we can better see the effect of only the satellite orientations. All fits discussed in this section are fits to central position, satellite orientation correlation functions. We do note that in the 2-halo regime, it is expected that the overall alignment correlation functions are dominated by the impact of central-central alignment only \citep{Schneider:2010cg, Fortuna_2020}; this is an important caveat to bear in mind in the following discussions in regards to the impact of our model's ability or inability to fit central-satellite correlations well in the 2-halo regime.

\begin{figure*}
    \centering
    \includegraphics[width=2\columnwidth]{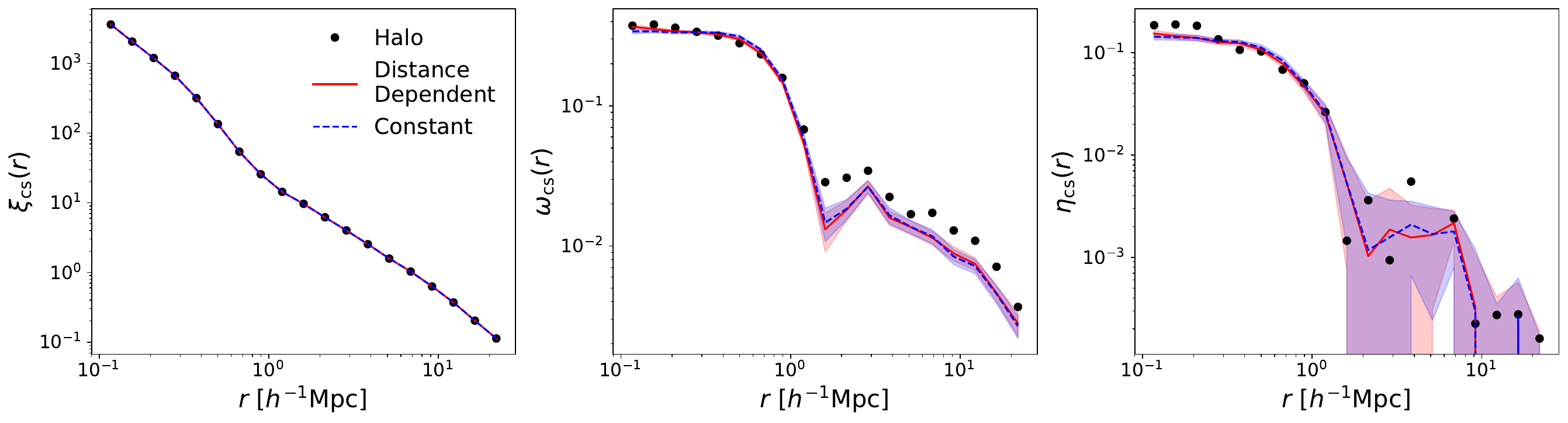}
    \caption{{\bf Left:} $\xi_{\rm cs}$ correlation function (central galaxy position correlated with satellite galaxy position). Galaxy correlation functions are shown in the solid red and dashed blue lines, with the corresponding halo correlation function shown in black dots. Here, all three share the same $\xi$ (position--position) correlation function as the only difference in the model is the alignment. {\bf Middle:} $\omega_{\rm cs}$ correlation function (central galaxy position and satellite galaxy orientation). Here, we see a slight difference between $\omega^{\rm g}_{\rm cs}$ (the galaxy correlation functions) and $\omega^{\rm h}_{\rm cs}$ (the corresponding halo correlation function) with the distance-dependent alignment (solid red line) performing slightly better at small scales than the constant alignment strength (shown as a dashed blue line). {\bf Right:} $\eta_{\rm cs}$ correlation function (central galaxy orientation and satellite galaxy orientation). In all panels (and in figures below), the shaded regions show the 1-sigma confidence interval of each model.}
    \label{fig:cen_sat_correlations}
\end{figure*}

As shown in central plot of Figure \ref{fig:cen_sat_correlations}, and discussed in more detail in Section \ref{sec:subhalo_ia}, the distance-dependent and constant radial alignment strengths perform comparably. The correlation function from the distance-dependent radial alignment may match that of its halos slightly better, but adds an extra degree of complexity. We also see that the correlation functions match much better in the 1-halo regime (here, defined as anything within 1 Mpc/$h$) than overall.

\begin{figure*}[!ht]
    \centering
    \includegraphics[width=2\columnwidth]{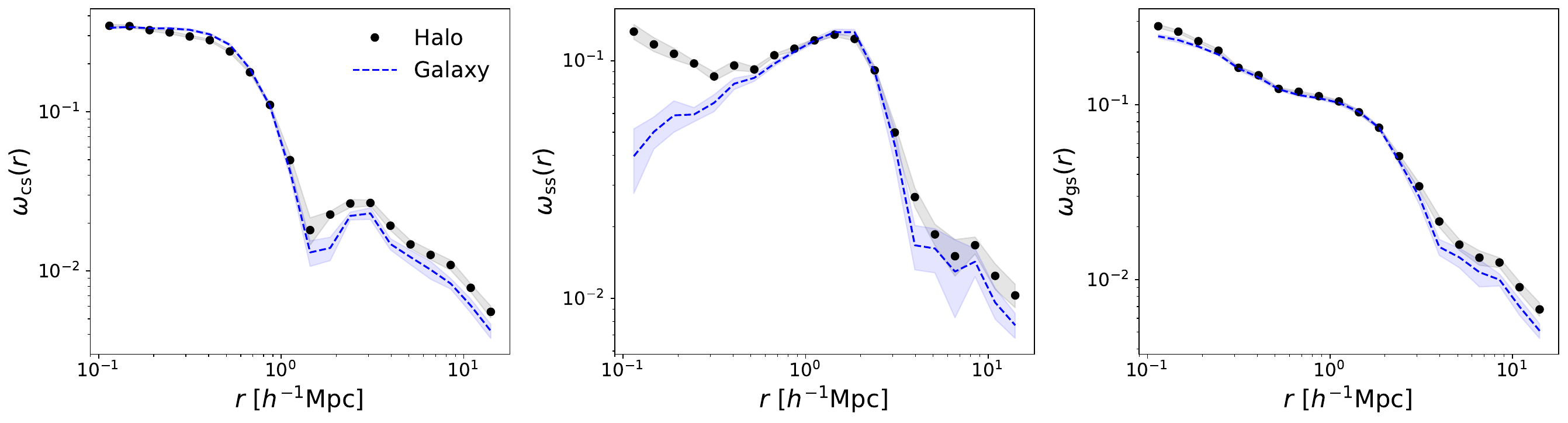}
    \caption{The position--orientation correlations for central--satellite (left; $\omega_{\rm cs}$), satellite--satellite (middle; $\omega_{\rm ss}$), and all galaxy--satellite (right; $\omega_{\rm gs}$) pairs. All correlations shown are using a constant radial alignment model. {\bf Left:} $\omega_{\rm cs}$ function; an identical configuration to the constant radial alignment series in the middle panel of Figure \ref{fig:cen_sat_correlations}. Here we see good agreement to the 1-halo regime with slight underestimation in the 2-halo regime. {\bf Middle:} $\omega_{\rm ss}$ correlation function. Here we see a generally good agreement except at small scales. {\bf Right:} $\omega_{\rm gs}$ correlation function. We see that, although $\omega_{\rm ss}$ performs poorly at small scales, the $\omega_{\rm cs}$ portion dominates making the overall correlation a good fit.}
    \label{fig:sat-sat}
\end{figure*}

For the fits shown in Figure \ref{fig:cen_sat_correlations}, we determined parameters by fitting the 1-halo portion of the $\omega_{\rm cs}$ correlation functions from the galaxies to those of the halos using MCMC. We discuss our decision to fit solely to the 1-halo regime in Section \ref{sec:1_halo_fit}, and we note that we use this type of fitting solely to show the flexibility of the model, as the radial alignment model is intended for situations where subhalo orientations are not known. We find the best constant radial alignment strength to be $\mu_{\rm sat}=0.826$, and the best distance-dependent alignment strength, following equation \eqref{eq:galaxy_alignment_strength}, to have $\text{a} = 0.804$ and $\gamma = -0.04$.

While radial alignment is intended for a situation in which no subhalo orientations are available, we do have subhalo orienations and can use them for comparison purposes now. Comparing to the subhalo correlation functions, we see in Figure \ref{fig:cen_sat_correlations} that the distance-dependent alignment strength produces correlation functions that match within $40\%$ of the halo correlation functions overall (much better on average), and well within $15\%$ in the 1-halo regime (much better on average: about $6\%$). Meanwhile, the constant alignment strength produces results with very similar values. However, in the 1-halo regime, the average percent difference between the constant model and the halos is slightly worse: about $8\%$ off on average. A closer examination of the relative error can be seen in Section \ref{sec:1_halo_fit} in the context of fitting to the 1-halo and 2-halo regimes; both the distance-dependent and constant models here closely match those seen in the 1-halo fit model described there. While there may be a slight preference for a distance-dependent radial alignment, the two methods perform comparably. Certainly, constant radial alignment provides a simpler model and may be sufficient for many needs.

In Figure \ref{fig:cen_sat_correlations} we restricted our view to central--satellite correlations. We have done this because the radial alignment model inherently contains central--position, satellite--orientation information, but we noted earlier that the 2-halo correlation function can be decomposed into central--central, central--satellite, and satellite--satellite portions. In Figure \ref{fig:sat-sat} we see a similar decomposition of position--orientation correlation functions into central--satellite ($\omega_{\rm cs}$), satellite-satellite ($\omega_{\rm ss}$), and all galaxy--satellite ($\omega_{\rm gs}$) pairs. The ($\omega_{\rm cs}$) correlation is identical to the constant radial alignment portion of the central panel of Figure \ref{fig:cen_sat_correlations}, but the other panels look at other contributions. As we have seen, ($\omega_{\rm cs}$) fits the 1-halo regime well and slightly underestimates the 2-halo regime. Meanwhile, the ($\omega_{\rm ss}$) portion performs reasonably well until we reach small scales. In the ($\omega_{\rm gs}$) panel, we see that, though ($\omega_{\rm ss}$) fails at small scales, the ($\omega_{\rm cs}$) portion dominates, making the overall agreement good.

\subsection{Fitting to the 1-halo and 2-halo Regimes}
\label{sec:1_halo_fit}
We have compared the galaxy correlation functions to those of their respective halos for both the full range, as well for the 1-halo regime alone, generally finding that our models fared better on small scales. Because of this, we restrict ourselves to the 1-halo regime when finding the best-fit parameters for the constant and distance-dependent radial alignment models. In this section, we further justify that decision by showing the limits of our model when considering satellite alignments. For both the constant radial alignment strength and the distance-dependent radial alignment strength cases, we ran MCMC chains and compared the galaxy correlation functions to the halo correlation functions in the 1-halo regime. To run these chains, and any time we use MCMC, we use the {\tt emcee} python package \citep{2013PASP..125..306F}.

Our attempts to fit to the correlation function across all scales led to a wide spread of $a$ and $\gamma$ values for equation \eqref{eq:galaxy_alignment_strength}, producing extreme values for $\mu$, and resulting in poor fits and no clear best-fit parameter values. Our attempts to fit to large-scale correlations alone (i.e., the 2-halo regime) resulted in similar issues. Meanwhile, when restricting attention to the small-scale correlations alone, our fitter enjoyed higher-accuracy success. These results, and the fact that we underestimate the 2-halo correlations when fitting to the 1-halo regime only, imply that there is extra environmental information needed to reproduce correlation functions between central position and satellite orientation in the 2-halo regime, and that a simple radial alignment does not fully capture subhalo orientations. This situation is not surprising, since our model only directly includes information from the 1-halo regime (e.g.\ satellite positions and (sub)halo orientations).

\begin{figure}[!ht]
    \centering
    \includegraphics[width=\columnwidth]{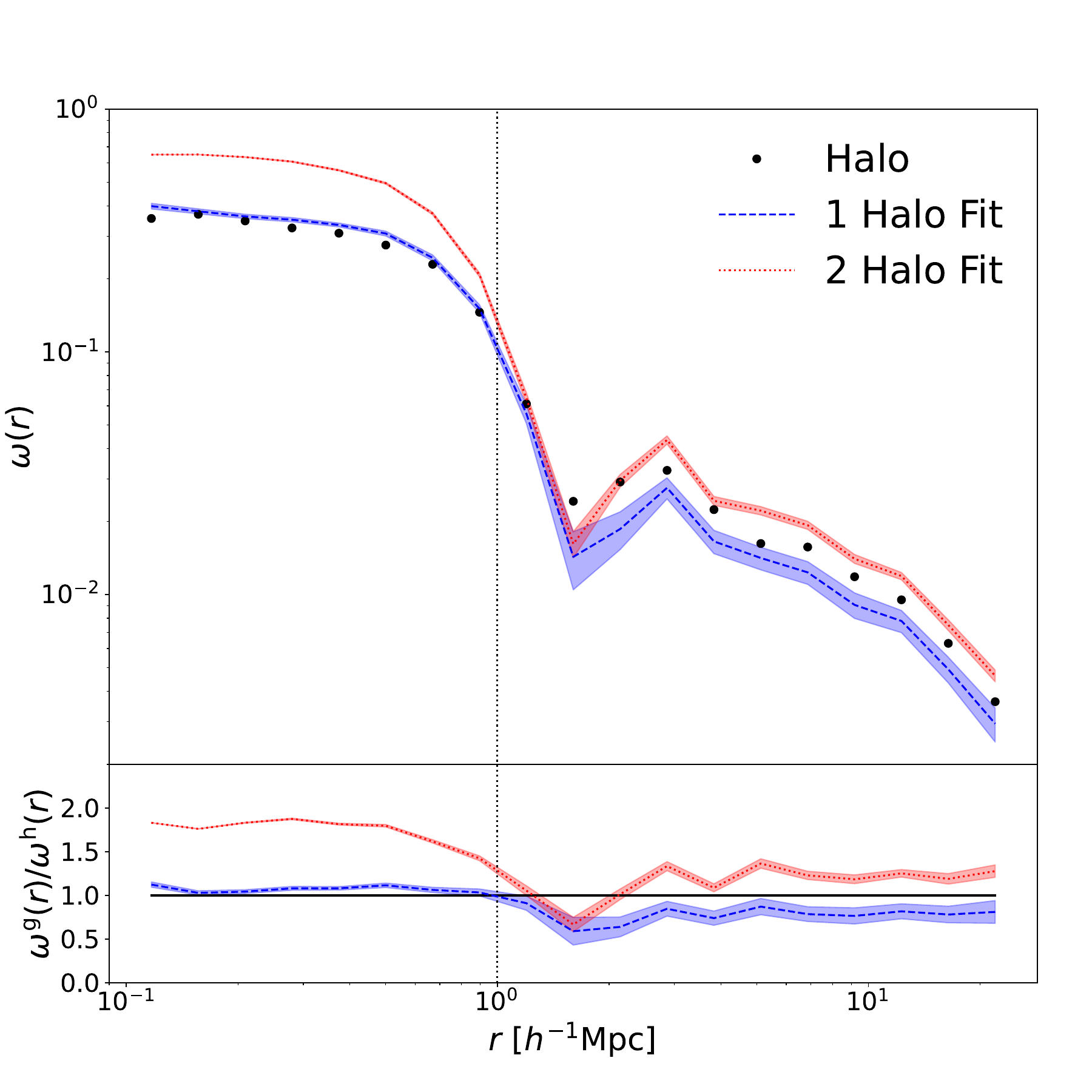}
    \caption{{\bf Top:} Comparison of the central-satellite position-orientation correlation function for halos, $\omega^{\rm h}_{\rm cs}$ (black points), with that of the galaxies, $\omega^{\rm g}_{\rm cs}$ (colored lines). Two galaxy series are shown, each one having been generated by an HOD model where the galaxy alignment strengths were fit using MCMC. {\bf Bottom:} The ratio of the galaxy correlation function over the halo correlation function ($\omega^{\rm g} / \omega^{\rm h}$). The series where the parameters were fit to the 1-halo part of the halo correlation functions fits closely in the 1-halo regime (agreeing within $15\%$), as expected, while failing to accurately represent the 2-halo regime (differing by up to $40\%$). Meanwhile, the HOD with alignment strengths obtained by fitting to the 2-halo regime fail to model the 1-halo signal (differing by up to $90\%$), and don't follow the 2-halo correlation function well (only agreeing within $40\%$) despite having been fit to that area.}
    \label{fig:one_halo_two_halo}
\end{figure}

In Figure \ref{fig:one_halo_two_halo}, we see a direct comparison of $\omega^{g}_{cs}$ (the central galaxy position, satellite galaxy orientation correlation function), as measured directly from simulated (sub)halos, and two different HOD models: one with alignment strengths found by fitting to the 1-halo regime only (labeled as the 1-halo fit model), and one defined by fitting to the 2-halo regime only (which we will call the 2-halo fit model). Overall, the 1-halo fit model performs better; it matches the 1-halo regime well and performs comparably to the 2-halo fit model in the 2-halo regime. Neither fit suggests that our model can accurately capture the full range of spatial scales, but looking at the 1-halo and 2-halo regimes separately, we see the strengths and weaknesses of each model in more detail. In the 1-halo range of scales, the 1-halo fit model achieves $\sim12\%$ accuracy at worst, while the 2-halo fit model only matches within $90\%$. In the 2-halo regime, the 1-halo fit model is only accurate at the $40\%$ level, but even the 2-halo fit model performs no better, differing by almost as much.

For a clearer picture, the $\chi^2$ per degree of freedom ($\chi^2_{\rm dof}$) (describing how well the galaxy correlation functions match those of their halos) of the 1-halo fit model is $5.2$, $8.6$, and $3.8$ for the full, 1-halo, and 2-halo regimes respectively, while the 2-halo fit model had values of $26160$, $55792$, and $10.8$. At this point, the only clear meaning of the 2-halo fit $\chi^2_{\rm dof}$ values is that fitting this model to the 2-halo regime does not work without extra information not yet captured by our model. We remind the reader that at this point, we still have made no attempt to refine the radial alignment model, but rather we have thus far only illustrated why we have chosen to fit our parameters to the 1-halo regime alone. In Section \ref{sec:subhalo_ia}, we turn attention to fitting distance-dependent radial alignment parameters and refining the goodness of fit.

We conclude this section by noting that our results imply a need for an additional modeling ingredient that captures the effect of environment beyond what is contained in the radial vector, similar to such ingredients that are now commonly used to capture assembly bias in HODs \citep[e.g.,][]{hearin_etal16_decorated_HODs,yuan_etal18_decorated_hod}. The nature of this correction may require appeal to some measure of large-scale density smoothed on some scale, and/or tidal information. We acknowledge this limitation, however, as demonstrated here, satellite alignments based on radial vectors are a simple, flexible, and effective model when simulated subhalos are unavailable. The development of a model with additional environmental dependence is beyond our current scope and is left as a topic for future work.

\subsection{Satellite Anisotropy}
\label{sec:sat_anisotropy}

In \S\ref{sec:radial_alignment}-\ref{sec:1_halo_fit}, we restricted consideration to HOD-style models of satellite alignments in which the spatial positions of satellites were assigned based on the positions of their respective subhalos. We know that such satellite anisotropy affects correlations \citep{samuroff_2020}, and for IA models of satellites based on the radial vector, anisotropy in satellite locations has an influence on the orientation correlation functions. In this section we study the effect of satellite isotropy and anisotropy on measurements of IA by comparing an isotropic phase space model with anisotropic satellite placement.

We model the anisotropic number density profile of satellites in halos as a triaxial NFW profile following \citet{Jing_2002} and \citet{Schneider_2012}.
\begin{equation}
\rho(R) = \frac{\rho_c}{\frac{R}{R_s}\left( 1+\frac{R}{R_s} \right)^2}
\end{equation}
where the relation between the Cartesian and elliptical coordinates is given by:
\begin{align}
x &= r\sin(\theta)\cos(\phi) = R\left(\frac{a}{c}\right)^{-\beta}\sin(\Theta)\cos(\Phi) \nonumber \\
y &= r\sin(\theta)\sin(\phi) = R\left(\frac{b}{c}\right)^{\beta}\sin(\Theta)\sin(\Phi) \nonumber \\
z &= r\cos(\theta) = R \cos(\Theta)
\end{align}
where $a$, $b$, and $c$ are the principle axis lengths of the host halo normalized such that $a=1$. The anisotropy parameter, $\beta$, controls the magnitude of the anisotropy; when $\beta=0$, satellites follow a spherically symmetric NFW profile; when $\beta=1$, satellites follow a triaxial NFW profile with axis ratios equal to that of the underlying dark matter halo; when $\beta>1$, the satellite distribution exhibits a greater degree of anisotropy than the underlying halo.

We compare three levels of anisotropy in the spatial distribution of satellites within their halos:

\begin{itemize}
    \item A subhalo position model, which places satellites at the locations of subhalos.
    \item A semi-isotropic model, which initially places galaxies using subhalo positions, but then randomly rotates each host halo's satellite population about the halo major axis. This model therefore preserves correlations involving the major axis but erases other anisotropies in the satellite distribution.
    \item An isotropic model, which places galaxies following a spherically symmetric NFW profile (as in \S\ref{sec:radial_alignment}-\ref{sec:1_halo_fit}.
\end{itemize}
We stress that in all three models sketched above, the only difference between the models is the level of anisotropy in the intra-halo positions of satellites. For the model of orientations explored in this section, here we assume that centrals are perfectly aligned with the halo shape, and for satellites we adopt a radial alignment model with constant strength of $0.85$.

\begin{figure*}[!ht]
    \centering
    \includegraphics[width=\textwidth]{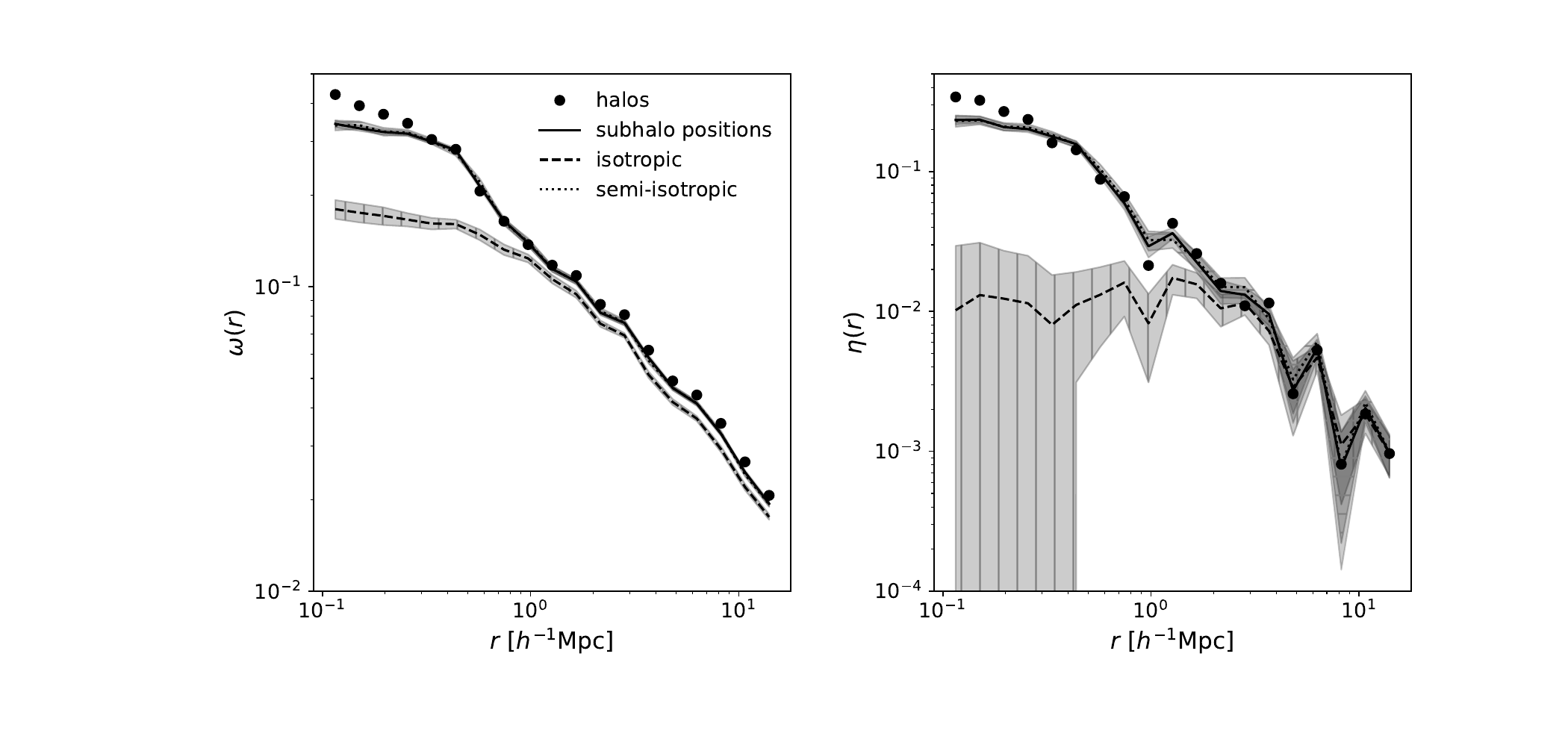}
    \caption{The effect of subhalo anisotropy on $\eta$ and $\omega$. Here we compare the correlations of all of the halos themselves (black points) to those of all of the galaxies (black lines), where the isotropy of galaxy placement is the only difference between the three galaxy correlation functions shown. The galaxy correlations shown are the average of fifty iterations (assigning position and alignment fifty times). While all three models align galaxies with respect to the radial vector, the subhalo position model uses subhalo positions to place galaxies while the isotropic model places the galaxies isotropically, and the semi-isotropic model uses the subhalo positions then rotates them around the halo major axis to preserve the radial distribution. Here we see that using the subhalo positions, even if we rotate them, produces a signal remarkably close to that of the subhalos. Placing galaxies isotropically appears to lose some information required to match the halo correlation functions with the parameters given.}
    \label{fig:anisotropy}
\end{figure*}

As shown in Figure \ref{fig:anisotropy}, both the subhalo position and semi-isotropic models allow the galaxy correlation functions to match those of the halos within $10\%$ for the $\omega$ correlation functions and within $30\%$ for the $\eta$ correlation function. Meanwhile, models with isotropic satellites predict much weaker correlations, especially in the 1-halo regime, resulting in a $30\%$ and a $60\%$ difference from the subhalo $\omega$ and $\eta$ correlation functions, respectively. The results illustrated in Figure \ref{fig:anisotropy} demonstrate that the level of anisotropy of satellite positions within the halo comprises an important degree of freedom for orientation correlation functions and that the major axis of the anisotropy contains most of the relevant information.

\subsection{Capturing Subhalo IA in Gravity-only Simulations}
\label{sec:subhalo_ia}

So far in this section, we have been primarily focused on the position--orientation correlation function; in this section we consider additional orientation correlation functions to study the flexibility of our satellite IA models. Figure \ref{fig:cen_sat_correlations} shows how three different correlation functions are impacted by satellite IA models with either constant or distance-dependent alignment strength. Each prediction was computed from 50 realizations of the model, and in all cases galaxy positions are placed at the center of (sub)halos. We remind the reader that the satellite fraction is $\sim25\%$ in the predicted galaxy population, and so our plots show central-to-satellite correlation functions of each type: the central position to satellite position correlation function, $\xi,$ the central position to satellite orientation correlation function, $\omega$, and the central orientation to satellite orientation correlation function,  $\eta$. We plot central-satellite cross-correlation functions to isolate the effect of satellite alignments on the IA signal.

The orientation correlation function $\eta$ is better fit in the 1-halo regime, again indicating that additional environmental information beyond simple radial alignment is needed in order to capture physically plausible satellite effects on IA in the 2-halo regime. As mentioned in Section \ref{sec:radial_alignment}, by fitting to the 1-halo regime, we found the best constant radial alignment strength to be $0.826$, and the best distance-dependent alignment strength to have $\text{a} = 0.804$ and $\gamma = -0.04$.

\begin{figure*}
    \includegraphics[height=\columnwidth]{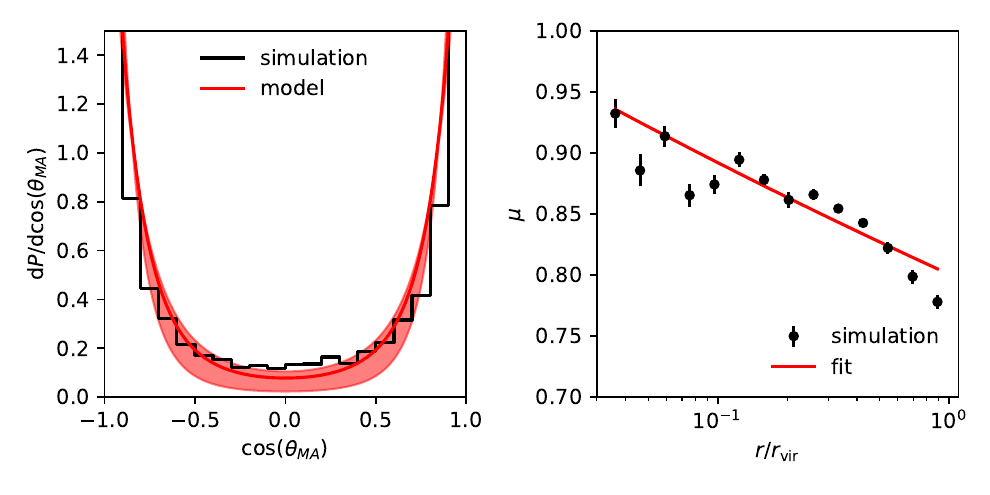}
    \caption{{\bf Left}: Differential distribution of (sub)halo radial misalignment angles for extant (sub)halos with a peak mass greater than $10^{12}~h^{-1}M_{\odot}$.  The best fit Dimroth-Watson distribution ($\mu = 0.83$) is shown as a red line.  The shaded region shows the variation in the alignment strength for 80\% of subhalos (i.e. the alignment strength assigned to 80\% of the subhalos based on where they fall on the power law fit in the right panel). We see here that a single Dimroth-Watson distribution provides a good fit to all (sub)halo misalignment angles. {\bf Right:} Dependence of the alignment strength for subhalos as a function of radial position, scaled by the virial radius of the host halo. The black points are alignment strengths for bins of $r/r_{\rm vir}$ (with the alignment strength of each bin using the process shown in the left panel) with error bars given by bootstrapping with 50 bootstrap realizations.The red line is a clipped power law fit to these points. As opposed to the left plot where we fit a single Dimroth-Watson distribution to all (sub)halos, we see here that if we bin (sub)halos by distance and fit a Dimroth-Watson distribution to the misalignment angles in each bin, the behavior is approximately described by a power law.}
    \label{fig:misalignment_distribution}
\end{figure*}

Previously, we derived these values using MCMC, but as an alternate method, we can directly fit the Dimroth-Watson distribution to the PDF of misalignment angles in the simulation. The results of this fitting process are shown in Figure \ref{fig:misalignment_distribution}, which displays the best-fitting Dimroth-Watson curve to the distribution of misalignment angles of simulated subhalos. To study the distance-dependence of alignment strength, we bin the subhalos (or satellite galaxies in the case of comparing to results obtained via MCMC) by $r/r_{vir}$, and perform separate fits on each bin. In the right panel of Figure \ref{fig:misalignment_distribution}, we show a power law fit to the resulting relationship between $\mu$ and $r/r_{vir}$.

At first glance, both the model with distance-dependent alignment strength and that with the constant alignment strength appear to give equally valid results for the $\omega$ correlation function, as they match within $20\%$ overall and within $6\%$ and $8\%$, respectively, in the 1-halo regime. However, there are noticeable differences when the full $\chi^2_{\rm dof}$ is considered. Looking at the $\omega$ (ED) correlation function, the model with distance-dependent alignment strength has a $\chi^2_{\rm dof}$ of $4.3$ overall, $3.9$ in the 1-halo regime, and $4.0$ in the 2-halo regime while the model with constant alignment strength shows a $\chi^2_{\rm dof}$ of $6.3$, $8.2$, and $4.3$ for the overall, 1-halo, and 2-halo regimes respectively. 
Similarly, if we look at the $\eta$ (EE) correlation functions, we see that the distance-dependent alignment strength matches the halo correlation functions within $10\%$ overall and within $13\%$ in the 1-halo regime, while the constant alignment strength model only matches within $24\%$ overall and $15\%$ in the 1-halo regime.
From this, we see that there does seem to be minor improvement from using a distance-dependent alignment model, especially when we limit our investigation to the 1-halo regime where our model works best. However, we should note the small value for $\gamma$ found above ($\gamma = -0.04$), indicating this power law dependence is very close to a constant radial alignment. 

In the 2-halo regime, the radial alignment model fails to capture the environmental dependence of subhalo orientations. In order to capture physically plausible satellite contributions to the IA signal on large scales, improving the physical realism beyond a simple radial alignment model will be necessary. We will explore such modeling improvements in future work.

\section{Capturing IA with Realistic Complexity: testing the model with Illustris}
\label{sec:illustris}

In the previous sections, we studied how models of satellite IA are able to capture the orientation correlation functions of subhalos in gravity-only N-body simulations. We now turn attention in this section to the IA signal exhibited by galaxies in the IllustrisTNG hydrodynamical simulation. In so doing, we will study IA models of both central and satellite galaxies acting together in concert. For central IA, we use the shape of the parent halo as the reference vector with respect to which central galaxy orientations are (mis)aligned. For satellite IA, we consider two different types of model: one based on the orientation of subhalos, the second HOD-type model based on radial alignments of constant strength. The free parameters of our model modulate the strength of central (satellite) galaxies with respect to the parent (sub)halo, and for all our results we use {\tt emcee} to fit the parameters regulating the alignment strength.

Prior to fitting the galaxy alignment strengths, we first adjust the occupation model parameters such that the halo occupation of our model more closely matches that of IllustrisTNG300. That is, we ensure that halos of a given mass are populated, on average, with the same number of galaxies as halos of comparable mass in IllustrisTNG300. We used the Zheng07Cens and Zheng07Sats  for the central and satellite galaxy occupation models available in {\halotools}, which calculate the mean central and satellite occupations following equations (2) and (5) of \citet{Zheng_2007}. Table \ref{tab:Illustris_HOD} shows the occupation parameters found by fitting the central and satellite occupation functions used by Zheng07Cens and Zheng07Sats to the mean occupations seen in Illustris. Specifically, the parameters were fit on the region where the central occupation transitions from zero central galaxies, to one central galaxy. For simplicity, in our fits $\sigma_{\log{M}}$ (the width of the cutoff profile) and $\alpha$ (asymptotic slope at high halo mass) were held fixed at $0.26$ and $1.0$ respectively and not allowed to vary. As noted previously, the models built using {\halotools} tend to have roughly three central galaxies to every one satellite, giving a satellite fraction of roughly $25\%$ with the occupation parameters used. The Illustris satellite fraction varies between $30\%$ and $40\%$ satellite galaxies, depending on the stellar mass threshold chosen (with higher stellar mass thresholds yielding a lower percentage of satellite galaxies).

\begin{figure*}
    \centering
    \begin{subfigure}[b]{0.3\textwidth}
       \centering
       \includegraphics[width=\textwidth]{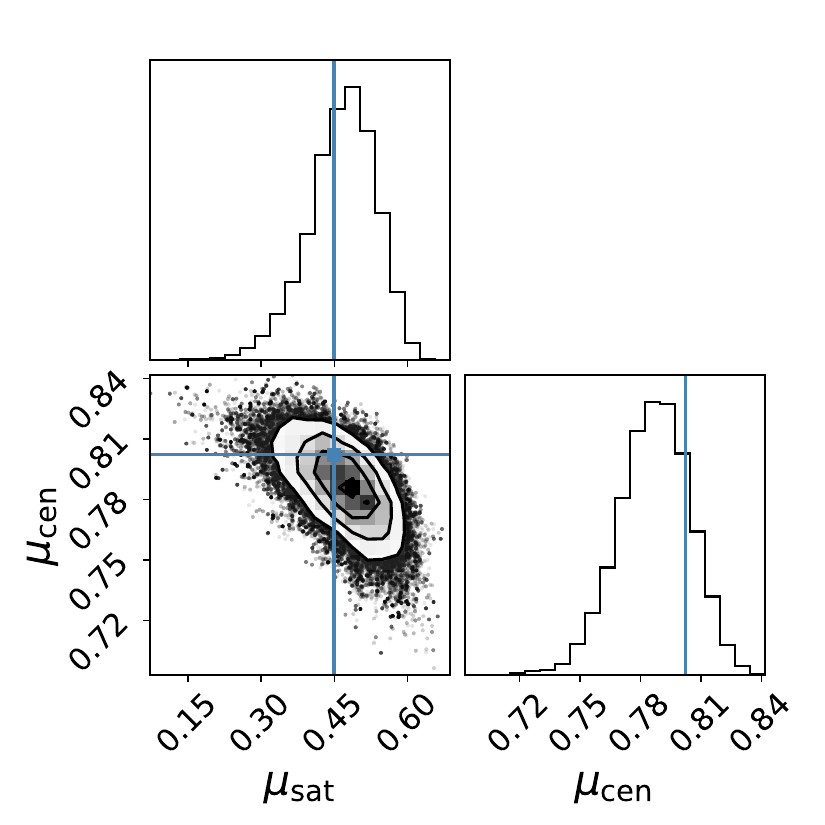}
       \caption*{$\log{M_*}\ge10.5$}
    \end{subfigure}
    \begin{subfigure}[b]{0.3\textwidth}
       \centering
       \includegraphics[width=\textwidth]{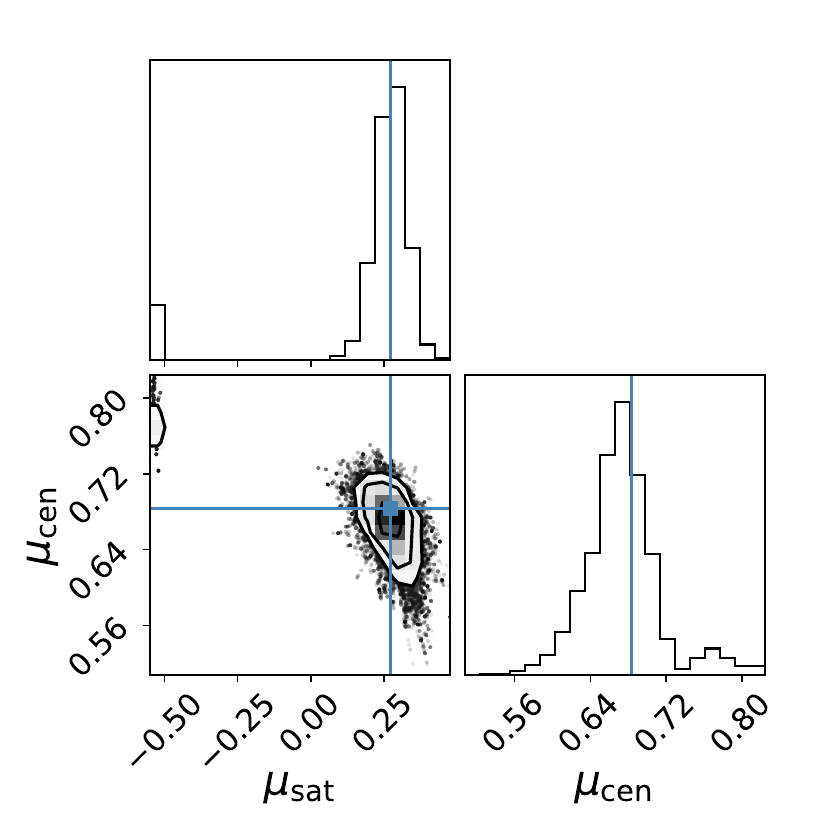}
       \caption*{$\log{M_*}\ge10.0$}
    \end{subfigure}
    \begin{subfigure}[b]{0.3\textwidth}
       \centering
       \includegraphics[width=\textwidth]{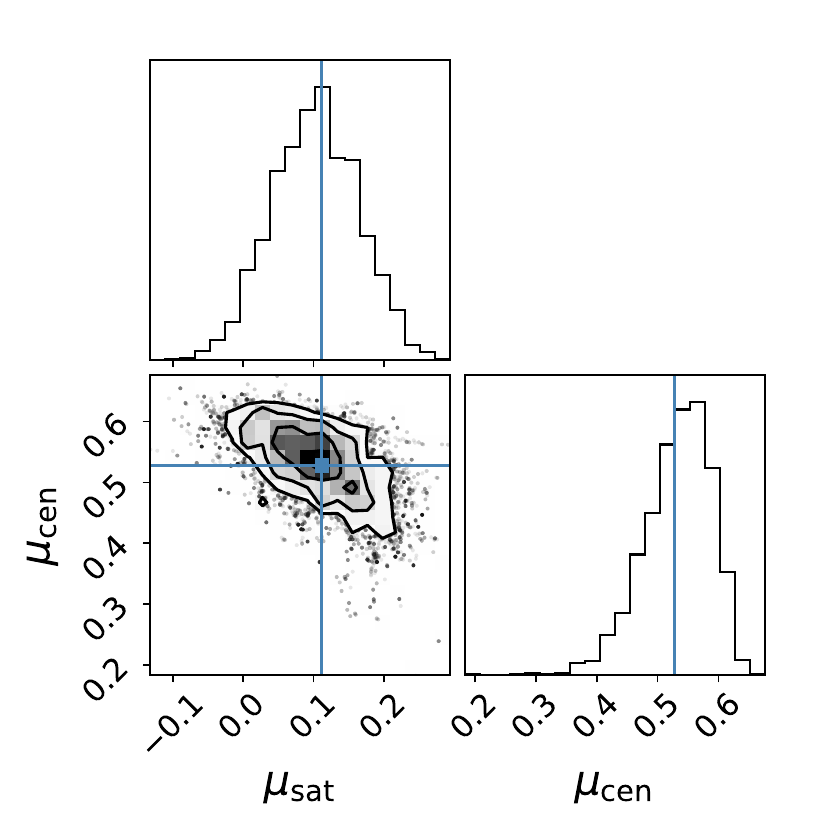}
       \caption*{$\log{M_*}\ge9.5$}
    \end{subfigure}
    \caption{The parameter values found from MCMC for central alignment strength ($\mu_{\rm cen}$) and satellite alignment strength ($\mu_{\rm sat}$) to let HOD models best fit Illustris correlation functions using a subhalo alignment model for the satellite galaxies. Each sample corresponds to a different mass cutoff for the halos included in the HOD model. These are the same mass cutoffs seen in Figure \ref{fig:three_panel_illustris}. In the models used to generate these corner plots, we used a central alignment for central galaxies and subhalo alignment for satellite galaxies. We did the same process using constant radial alignment for satellite galaxies, yielding similar corner plots not shown here. As a general trend, we see the alignment strengths increase with higher mass cutoffs. \textbf{Left:} Best parameters found were $\mu_{\rm cen} = 0.802$ and $\mu_{\rm sat} = 0.449$. \textbf{Middle:} Best parameters found were $\mu_{\rm cen} = 0.683$ and $\mu_{\rm sat} = 0.274$. \textbf{Right:} Best parameters found were $\mu_{\rm cen} = 0.528$ and $\mu_{\rm sat} = 0.112$.}
    \label{fig:Illustris_MCMC}
\end{figure*}

\begin{figure}
    \includegraphics[width=\columnwidth]{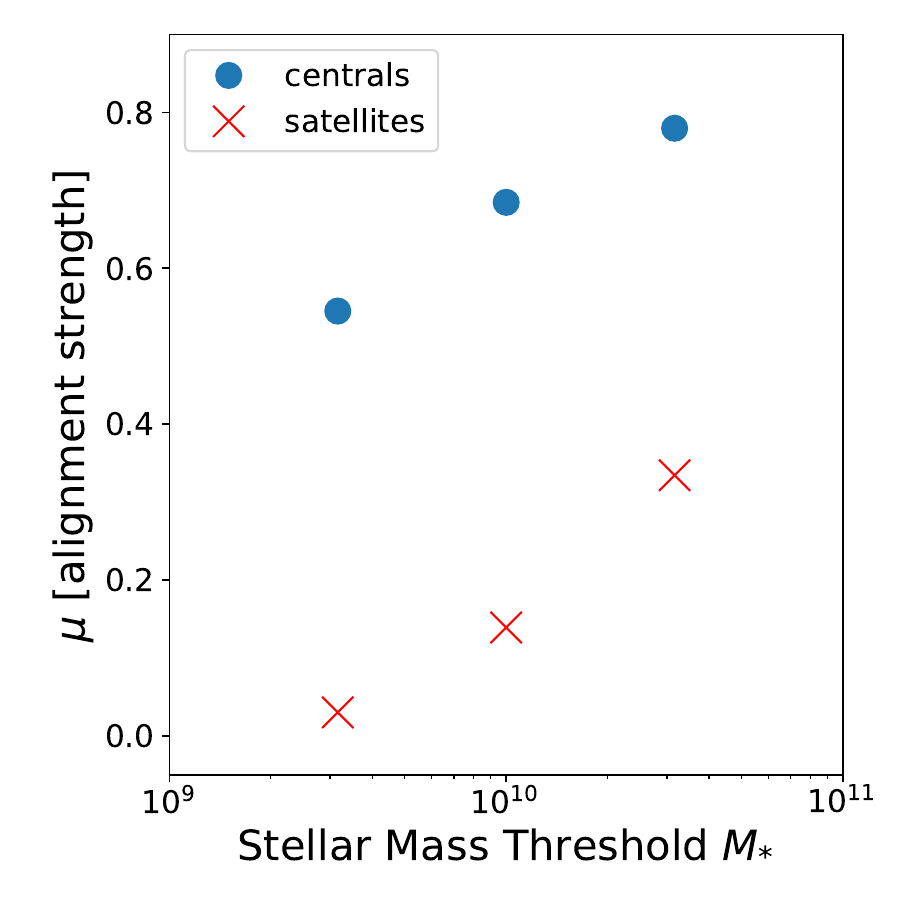}
    \caption{Alignment using the central alignment and subhalo alignment models for the central and satellite galaxies respectively. The three points shown for each type of galaxy correspond to samples one, two, and three. The samples are separated based on the minimum mass threshold for the galaxies included in the sample, shown on the x-axis.}
    \label{fig:alignment_trend}
\end{figure}

The corner plots in Figure \ref{fig:Illustris_MCMC} display the posteriors from our fits to the orientation correlation functions in Illustris. As indicated in the figure labels, we show results for fits to three separate galaxy samples, each defined by a different stellar mass cut. Sample 1 corresponds to $\log{(M_*)} \geq 10.5$, sample 2 to $\log{(M_*)} \geq 10$, and sample 3 to $\log{(M_*)} \geq 9.5$.  Two broad trends are apparent in Figure \ref{fig:Illustris_MCMC}. First, the central galaxies have a noticeably larger alignment strength than the satellite galaxies. Second, as the stellar mass threshold increases, so does the alignment strength of both populations. Both of the above trends can be more clearly seen in Figure \ref{fig:alignment_trend}, which shows how the best-fitting alignment strength $\mu$ changes with the stellar mass cut.

\begin{figure*}
    \includegraphics[width=2.2\columnwidth]{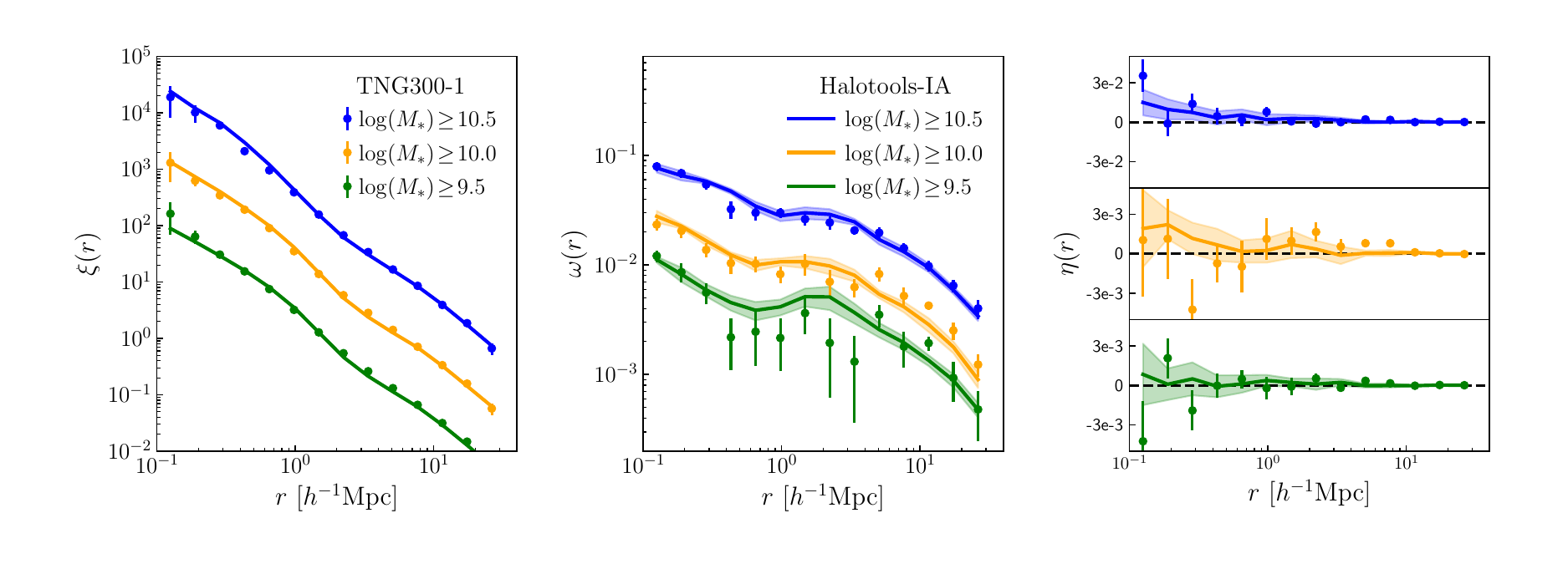}
    \caption{{\bf Left:} Two-point correlation functions offset by 1-dex for clarity {\bf Middle:} position--orientation correlation function {\bf Right:} orientation--orientation correlation function.  In each panel, the points with error bars are measurements made on the IllustrisTNG300-1 simulation with error bars estimated using jackknife re-sampling of the box. The lines with shaded regions are halo model predictions made by populating a DMO simulation with mock galaxies where the shaded region shows the variation from random realizations of the model. We used a constant radial satellite alignment model in each case to showcase the flexibility of the model in cases where subhalo orientations are not available. The three colors are for three stellar mass threshold samples.}
    \label{fig:three_panel_illustris}
\end{figure*}

As seen in Figure \ref{fig:three_panel_illustris}, our best-fitting HOD model using constant radial satellite alignment generates correlation functions that agree reasonably well with IllustrisTNG. It is by now well established that an HOD model more complex than the one we use here is required for precision fits to the spatial clustering of hydro-simulated galaxies \citep[e.g.,][]{beltz_mohrmann_etal20,hadzhiyska_etal20}. For the $\xi$ correlations functions, the HOD models for samples one, two, and three produce correlations within $14.0\%$, $9.5\%$, and $12.0\%$ respectively to their corresponding counterpart from the Illustris data. This level of agreement is a sufficient baseline for purposes of broadly assessing the accuracy of our alignment strength models.

The center and right panels of Figure \ref{fig:three_panel_illustris} display the level of success of our best-fitting IA model to capture the position-orientation and orientation-orientation correlation functions, respectively. Table \ref{tab:illustris_chi_squared} shows the $\chi^2_{\rm dof}$ values for each of these fits. For the satellite IA, we find that a model based on radial alignment performs comparably well to a model based on subhalo alignment, which is encouraging for the prospects of HOD-type approaches to capturing orientation correlation functions.

\begin{table}
    \centering
    \begin{tabular}{ccccc}
         \hline
         Satellite & $\omega$ & $\eta$ & Central & Satellite \\
         Alignment & & & Alignment & Alignment \\
         Type & & & Strength & Strength \\ 
         \hline \hline
         Radial ($\log{M_*}\ge10.5$)  & 1.9 & 1.3 & 0.780 & 0.335 \\
         Radial ($\log{M_*}\ge10.0$) & 2.1 & 1.2 & 0.685 & 0.139 \\
         Radial ($\log{M_*}\ge9.5$) & 1.4 & 0.5 & 0.545 & 0.030 \\
         \hline \hline
         Subhalo ($\log{M_*}\ge10.5$) & 1.8 & 1.2 & 0.802 & 0.449 \\
         Subhalo ($\log{M_*}\ge10.0$) & 1.7 & 1.1 & 0.683 & 0.274 \\
         Subhalo ($\log{M_*}\ge9.5$) & 1.8 & 0.7 & 0.528 & 0.112 \\
         \hline
    \end{tabular}
    \caption{$\chi^2_{\rm dof}$ values for the central-satellite $\omega$ (position-orientation) and $\eta$ (orientation-orientation) correlation functions comparing the functions from our HOD models with those from Illustris. Also included are the alignment strengths that best fit our HOD model to IllustrisTNG. As before, these are the same mass cutoffs seen in Figure \ref{fig:three_panel_illustris}.}
    \label{tab:illustris_chi_squared}
\end{table}

\section{Discussion \& Conclusions}
\label{sec:conclusions}
In this work, we have studied forward modeling techniques that enable predictions of large-scale structure that include intrinsic alignments in galaxy shape. Our methodology is fully simulation-based: when predicting galaxy correlation functions, we first create a synthetic universe of galaxies by populating simulated halos, and then we compute summary statistics with the same point estimators used in the corresponding observational measurements. Our prediction pipeline is computationally efficient, and it has the practical capability to run likelihood analyses in the deeply nonlinear regime with MCMC sampling algorithms. We have demonstrated the effectiveness of our model by reproducing IA signals exhibited by subhalos in gravity-only simulations, as well as by galaxies in hydrodynamical simulations such as IllustrisTNG.

Our simulation-based methodology is flexible, and it is able to leverage information from subhalos when available in a high-resolution simulation, or alternatively to populate catalogs of host halos in lower-resolution simulations. When subhalo information is available in the underlying simulation, we can generate more physically realistic predictions by using the orientation of subhalos as the reference vector for satellite galaxies. When using simulations that do not include subhalo orientations (as is often the case due to the computational demands of survey-scale simulations), a radial satellite alignment model (in which satellites preferentially align with respect to the radial vector between central and satellite) can approximate reasonably well the IA signal exhibited by hydro-simulated galaxies in the 1-halo regime. 

In detail, we find that a distance-dependent alignment strength, following equation \eqref{eq:galaxy_alignment_strength}, may provide a slightly closer match to the 1-halo term signal seen in the correlation functions of either gravity-only subhalos or hydro-simulated galaxies. However, even a satellite model with a simple constant radial alignment strength produces a reasonable correlation function on small scales; looking specifically at $\omega^{\rm g}_{\rm cs}$, we see that the distance-dependent alignment strength with the radial alignment model gives a $\chi^2_{\rm dof}$ of 4.3 overall, while the constant alignment strength gives 6.3.

As shown in Figure \ref{fig:cen_sat_correlations}, no form of radial alignment model is able to faithfully reproduce the central--satellite correlation functions in the 2-halo regime; this implies the need for the development of a new model of galaxy alignments that more flexibly incorporates environmental information, analogous to ongoing efforts to capture assembly-biased halo occupation statistics \citep[e.g.,][]{contreras_etal21_shame_assembias,yuan_etal22_abacushod,lange_etal23_s8_boss_rsd_lensing}. Reference vectors based in part on the tidal field are a promising way to incorporate such environmental effects \citep{Harnois-Deraps_2022}.

We find that the anisotropic spatial distribution of satellites within their host halos plays an important role in the strength of orientation correlation functions. Figure \ref{fig:anisotropy} shows that models based on the anisotropic positions of subhalos substantially improve upon the realism of the predictions. Meanwhile, models based on a spherically symmetric NFW profile perform poorly in the 1-halo regime. These results imply the need for IA models with improved sophistication in the intra-halo phase space distributions of satellites, echoing the well-established need for such ingredients to capture ordinary two-point clustering \citep[e.g.,][]{orsi_angulo_2018,hadzhiyska_etal23}.

One of our main goals was to test how well the IA signal in IllustrisTNG can be mimicked with a simple HOD-type model of galaxy orientations. In our tests, we used MCMC to fit our models of central and satellite alignment strength to the orientation correlation functions of IllustrisTNG galaxies, repeating this analysis separately for samples defined by three different stellar mass cuts. Figure \ref{fig:three_panel_illustris} summarizes the results of this analysis, and Table \ref{tab:illustris_chi_squared} records the alignment strengths of our best-fitting models. Subhalo-based models are particularly effective in their ability to generate a mock galaxy catalog with correlation functions similar to IllustrisTNG, attaining $\chi^2_{\rm dof}$ close to unity for every sample and correlation function considered. Satellite models based on constant radial alignment perform comparably well to the subhalo models, despite their relative simplicity. In future work, we will generalize the HOD-type methodology used here to incorporate joint correlations between IA strength and galaxy brightness, color, and redshift as seen in simulations \citep{Zjupa_etal22_tt_tng} and observations \citep{samuroff_etal23_des_ia_lum_color}. We will also work towards adding more sophisticated alignment models that take into account more detailed sets of halo properties and other information, including redshift, in order to produce more robust results \citep{xu2023physical}.

In a full forward model of galaxy correlation functions $\xi$, $\omega,$ and $\eta$, additional ingredients for the probabilistic distribution of galaxy shapes will be needed in addition to the alignment distribution model studied here. Once the model is able to accommodate full shape information, we expect that the results here would show much better agreement corresponding to the expanded functionality of our alignment models. Our mock-population framework lays the groundwork for carrying out Bayesian inference with simulation-based forward models of galaxy shape correlation functions. Our simulation-based approach simplifies the effort required to model systematic effects that can be challenging to capture analytically, such as galaxy assembly bias, halo exclusion in the transition between 1- and 2-halo regimes \citep{vandenbosch_etal13_clf,garcia_rozo_2019_halo_exclusion}, and beyond-linear halo bias \citep{mahony_2022,mead_2021}. Forward models based on mock-population methods also extend naturally to predict higher-order summary statistics, which can provide additional information not contained in the two-point functions studied here \citep{harnois_eraps_etal21_peaks_desy1}. With the additional effort outlined above, our models could be used to derive posteriors on the true intrinsic alignment strength of galaxies from observations on highly nonlinear scales, and to supply priors on the IA nuisance parameters used in cosmological analyses of larger scales. 

\section*{Acknowledgements}

This paper has undergone internal review in the LSST Dark Energy Science Collaboration. The internal reviewers were Joachim Harnois-D\'{e}raps, Eve Kovacs, and Contance Mahony; we thank them for helpful comments and discussion.

Author contributions: 
NVA wrote code, conducted the primary analysis, and wrote the main body of the text. DC conceived of and initiated the project and contributed significantly to the initial text. JB participated in initial project planning, contributed significant guidance and mentorship during the code creation and analysis stages, and authored portions of the text. CDL contributed significant guidance and mentorship during the code creation and analysis stages, as well as writing and revising portions of the draft. FL contributed guidance on code, model, and statistical aspects of the analysis. APH contributed to code development of the alignment model and summary statistics computations, and helped develop and revise the draft. RM provided significant mentorship and guidance during the project's initial stages and provided input on the paper's near-complete version.

NVA and JB are supported by NSF Award AST-2206563, the Roman Research and Support Participation program under NASA grant 80NSSC24K0088, and the US DOE under grant DE-SC0024787 and the SCGSR program, administered by ORISE which is managed by ORAU under contract number DE-SC0014664. RM was supported in part by a grant from the Simons Foundation (Simons Investigator in Astrophysics, Award ID 620789) and in part by the US Department of Energy grant DE-SC0010118.

This work was completed in part using the Discovery cluster, supported by Northeastern University’s Research Computing team. This research used resources of the National Energy Research Scientific Computing Center (NERSC), a U.S. Department of Energy Office of Science User Facility located at Lawrence Berkeley National Laboratory, operated under Contract No. DE-AC02-05CH11231.

The DESC acknowledges ongoing support from the Institut National de 
Physique Nucl\'eaire et de Physique des Particules in France; the 
Science \& Technology Facilities Council in the United Kingdom; and the
Department of Energy, the National Science Foundation, and the LSST 
Corporation in the United States.  DESC uses resources of the IN2P3 
Computing Center (CC-IN2P3--Lyon/Villeurbanne - France) funded by the 
Centre National de la Recherche Scientifique; the National Energy 
Research Scientific Computing Center, a DOE Office of Science User 
Facility supported by the Office of Science of the U.S.\ Department of
Energy under Contract No.\ DE-AC02-05CH11231; STFC DiRAC HPC Facilities, 
funded by UK BEIS National E-infrastructure capital grants; and the UK 
particle physics grid, supported by the GridPP Collaboration.  This 
work was performed in part under DOE Contract DE-AC02-76SF00515.

This work made use of Astropy:\footnote{http://www.astropy.org} a community-developed core Python package and an ecosystem of tools and resources for astronomy \citep{astropy:2013, astropy:2018, astropy:2022}.

\appendix

\section{Galaxy \& Halo Shapes}
\label{appendix:shapes}

We model (sub)halos and galaxies as 3-dimensional ellipsoids.  The shape and orientation of (sub)halos/galaxies may then be characterized by calculating the reduced inertia tensor for the particle distribution.  We define the reduced inertia tensor as:
\begin{equation}
\label{eq:reduced_inertia_tensor}
\tilde{\bf I}_{ij} = \frac{\sum m_n \frac{x_{ni} x_{nj}}{r_{n}^2}}{\sum m_n}
\end{equation}
where
\begin{equation}
r_{n}^2 = \sum x_{ni}^2
\end{equation}
is the distance between the center of mass and the $n^{\rm th}$ particle in the system.  The reduced inertia tensor applies more weight to particles that are near the center of the object, reducing the sensitivity to loosely bound particles present in the outer regions of the object.  

The triaxial shape of the (sub)halo/galaxy is specified by the eigenvalues of the inertia tensor, $\lambda_a > \lambda_b > \lambda_c$, where the half-lengths of the principle axis of the ellipsoid are given by $a=\sqrt{\lambda_a}$, $b=\sqrt{\lambda_b}$, $c=\sqrt{\lambda_c}$, ($a \geq b \geq c$).  The orientation of this ellipsoid is then specified by the eigenvectors, $\hat{e}_a$, $\hat{e}_b$, and $\hat{e}_c$.  

The radial weighting applied in equation \eqref{eq:reduced_inertia_tensor} tends to make (sub)halos/galaxies appear more spherical.  To alleviate this, we iteratively calculate the reduced inertia tensor for (sub)halos/galaxies.  After the initial calculation of the $\tilde{\bf I}$, the particle distribution is rotated so that $\hat{e}_a$, $\hat{e}_b$, and $\hat{e}_c$ are aligned with the x-, y-, x-axis.  The radial distance, $r_{n}^2$ in equation \eqref{eq:reduced_inertia_tensor}, is replaced with the elliptical radial distance:
\begin{equation}
r_{n}^2 = \left(\frac{x_n}{a}\right)^2 +
\left(\frac{y_n}{b}\right)^2 +
\left(\frac{z_n}{c}\right)^2
\end{equation}
This process is repeated until the eigenvalues change by less than 1\% between iterations.

We implicitly model (sub)halos/galaxies as homologous triaxial ellipsoids.  Note the caveats here.  halos contain substructure.  galaxies may contain multiple components.  galaxies are observed to have twisting isophotes.  By modeling galaxies as homologous ellipsoids, we can specify a single shape and orientation, vastly simplifying the modeling.

\section{Fitting Functions}
\label{appendix:fitting_functions}

In Figures \ref{fig:degrading_central_alignments} and \ref{fig:degrading_satellite_alignments}, we fit a smooth function to our measurements of simulated (subhalo) $\eta$ (EE) measurements.  We primarily do this to reduce the noise in the measurement at any given radius in order to make the figures more clear.  We present our fitting function here for completeness.
\begin{equation}
\eta(r) = A_1\exp\left(-(r/R_1)^{\gamma} \right) + \frac{A_2\exp\left( -(R_2/r) \right)}{1.0+(r/R_2)^{\alpha}}
\end{equation}
\begin{align}
\omega(r) = &A_1\exp\left(-(r/R_1)^{\gamma} \right) + \frac{A_2\exp\left( -(R_2/r) \right)}{1.0+(r/R_2)^{\alpha}} \\
 &+ \frac{A_3}{1.0+(r/R_3)^{\beta}}
\end{align}

\section{IllustrisTNG HOD}
\label{appendix:tnghod}
As shown in \S\ref{sec:illustris}, the alignment correlation functions  simultaneously depend on the misalignment distributions as well as the underlying halo occupation statistics. Thus in order to test the efficacy of our misalignment modeling  using IllustrisTNG, it's important to first establish an accurate baseline HOD model. In Table \ref{tab:Illustris_HOD}, we see the occupation model parameters used. These values were found by fitting the occupation parameters in the Zheng07Cens and Zheng07Sats occupation models to fit the occupation distribution of our HOD models with that of Illustris. Specifically, we fit the parameters in the region where the central occupation transitions from zero to one. Note that the values $\sigma_{\log{M}}$ and $\alpha$ have been fixed for simplicity, only allowing the other three parameters to vary.

\begin{table}
\centering
 \begin{tabular}{c c c c c c} 
 \hline
$\log(M_{\rm thresh})$ & $\log(M_{\rm min})$ & $\sigma_{\log{M}}$ & $\log(M_0)$ & $\log(M_1)$ & $\alpha$ \\ [0.5ex] 
 \hline\hline
 9.0 & 11.37 & 0.26 & 11.55 & 12.35 & 1.0 \\ 
 \hline
 9.5 & 11.61 & 0.26 & 11.8 & 12.6 & 1.0 \\
 \hline
 10.0 & 11.93 & 0.26 & 12.05 & 12.85 & 1.0 \\
 \hline
 10.5 & 12.54 & 0.26 & 12.68 & 13.48 & 1.0 \\
 \hline
\end{tabular}
\caption{HOD parameters for TNG300-1}
\label{tab:Illustris_HOD}
\end{table}

\section{Covariance}
\label{appendix:covariance}

Due to the probabilistic nature of populating an HOD model and aligning with a Dimroth-Watson distribution (discussed later in Section \ref{sec:model_misalignments}), there is stochasticity in both the data vector and the model. We have two different ways in which we select our covariance, depending on our data vector.

\begin{figure}[!ht]
    \centering
    \includegraphics[width=\columnwidth]{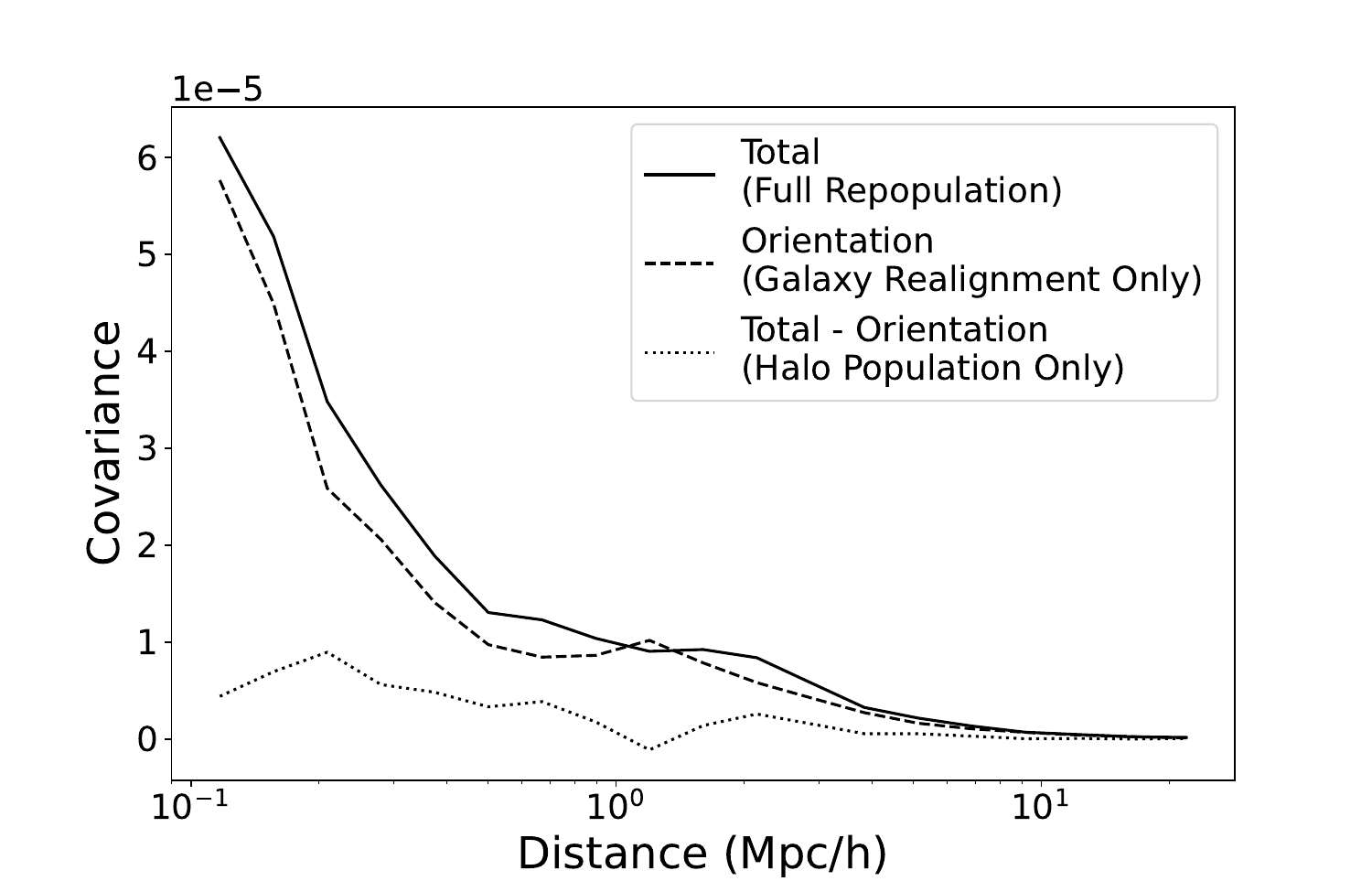}
    \caption{The diagonals of the covariance matrices generated from measuring the ed correlations of 100 instances of mock galaxy catalogs. The curve for total covariance comes from catalogs where we performed a full repopulation (i.e. no random seed was used and each time, halos were selected and populated randomly before assigning galaxy orientation). The orientation curve was generated in a similar fashion, but instead of allowing a full repopulation, we used a fixed seed for the halo population (i.e. the same halos were chosen each time with the same galaxy populations) meaning the only source of variance was the galaxy alignments. The final curve shows the difference between these two, Total - Orientation, representing the covariance that comes from just the repopulation of the halos, ignoring the galaxy alignments. Here we see that the majority of the covariance comes from realigning the galaxies.}
    \label{fig:covariance}
\end{figure}

In Section \ref{sec:no_subhalos}, we fit our satellite galaxy alignment strengths to find an alignment strength for the radial alignment such that the  central position, satellite orientation correlation functions for the galaxies match as closely as possible to those of the halos (keeping central galaxies perfectly aligned with their host halos for simplicity). However, because there is an element of stochasticity both in the population of the dark matter halos with galaxies, and in the realignment of the galaxies, both our data vector and our model have covariance. To account for this, we look at the contribution to the covariance of repopulating halos versus that of assigning galaxy alignments. As seen in Figure \ref{fig:covariance}, we find that most of the covariance comes from aligning galaxies. As such, we use a fixed seed for the galaxy population step in our MCMC, meaning the same dark matter halos are populated on each run, keeping the galaxy alignment step the only stochastic process. For our covariance, we use the covariance of $\omega^{\rm g}_{\rm cs}$ (the central galaxy position, satellite galaxy orientation correlation function) from 100 instances of realigning galaxies at that seed, for a given alignment strength (see Section \ref{sec:no_subhalos} for more detail). We use this covariance in the MCMC, after which a new alignment strength is determined, and the process repeats (generating a new covariance from 100 new iterations at the new alignment strength) until the alignment strength resulting from the MCMC does not change appreciably.

In Section \ref{sec:illustris}, rather than fitting to the halo orientations, we found both central and satellite galaxy alignment strengths by fitting $\omega^{\rm g}_{\rm gg}$ (the gg subscript now referring to all galaxies) correlation functions to those of IllustrisTNG300. In this case, since we had a fixed data vector, we used jackknife resampling of the Illustris data, splitting the simulation box into 5x5x5 jackknife regions to determine our covariance. It is important to note that we multiplied this jackknife covariance by the Hartlap factor, shown in equation \eqref{eq:hartlap} (taken from equation 17 in \citet{Hartlap_2007}) 

\begin{equation}
    \label{eq:hartlap}
    \hat{C}^{-1} = \frac{n - p - 2}{n - 1} \hat{C}^{-1}_*,
\end{equation}

where $\hat{C}^{-1}$ is the adjusted inverse covariance, $\hat{C}^{-1}_*$ is the inverse covariance directly from jackknife resampling, $n$ is the number of independent observations (i.e. the number of jackknife regions), and $p$ is the number of elements in the data vector (i.e. the number of bins). This helps approximately debias our estimation of the original inverse jackknife covariance.

\paragraph{Data availability}
We used the Bolshoi-Planck catalog and Multidark, both available through halotools and online at \href{https://www.cosmosim.org/}{https://www.cosmosim.org/}. The IllustrisTNG data we used is publicly available at \href{https://www.tng-project.org/}{https://www.tng-project.org/}.

\vspace{2 mm}

\bibliographystyle{mnras}
\bibliography{bib}
\end{document}